# MorphoITH: A Framework for Deconvolving Intra-Tumor Heterogeneity Using Tissue Morphology

Aleksandra Weronika Nielsen[1], Hafez Eslami Manoochehri[1,2], Hua Zhong[1,3], Vandana Panwar[3], Vipul Jarmale[1], Jay Jasti[1], Mehrdad Nourani[2], Dinesh Rakheja[3], James Brugarolas[4,5], Payal Kapur[3,4,6,*], Satwik Rajaram[1,3,4,*]

[1]Lyda Hill Department of Bioinformatics, University of Texas Southwestern Medical Center, Dallas, TX, USA
[2]Department of Electrical and Computer Engineering, The University of Texas at Dallas, Richardson, TX, USA
[3]Department of Pathology, University of Texas Southwestern Medical Center, Dallas, TX, USA
[4]Kidney Cancer Program, Simmons Comprehensive Cancer Center, University of Texas Southwestern Medical Center, Dallas, TX, USA
[5]Department of Internal Medicine (Hematology-Oncology), University of Texas Southwestern Medical Center, Dallas, TX, USA
[6]Department of Urology, University of Texas Southwestern Medical Center at Dallas, Dallas, TX, USA
[*]Corresponding authors: Payal Kapur (Payal.Kapur@UTSouthwestern.edu), Satwik Rajaram (Satwik.Rajaram@UTSouthwestern.edu)

**ABSTRACT**

The ability of tumors to evolve and adapt by developing subclones in different genetic and epigenetic states is a major challenge in oncology. Traditional tools like multi-regional sequencing used to study tumor evolution and the resultant intra-tumor heterogeneity (ITH) are often impractical because of their resource-intensiveness and limited scalability. Here, we present MorphoITH, a novel framework that leverages histopathology slides to deconvolve molecular ITH through tissue morphology. MorphoITH integrates a self-supervised deep learning similarity measure to capture phenotypic variation across multiple dimensions (cytology, architecture, and microenvironment) with rigorous methods to eliminate spurious sources of variation. Using a prototype of ITH, clear cell renal cell carcinoma (ccRCC), we show that MorphoITH captures clinically-significant biological features, such as vascular architecture and nuclear grades. Furthermore, we find that MorphoITH recognizes differential biological states corresponding to subclonal changes in key driver genes (*BAP1*/*PBRM1*/*SETD2*). Finally, by applying MorphoITH to a multi-regional sequencing experiment, we postulate evolutionary trajectories that largely recapitulate genetic evolution. In summary, MorphoITH provides a scalable phenotypic lens that bridges the gap between histopathology and genomics, advancing precision oncology.

**INTRODUCTION**

Intra-tumor heterogeneity (ITH), the coexistence of neoplastic cells in distinct molecular and functional states within the same tumor, is a critical challenge for precision oncology[1,2]. Driven by cancer evolution, ITH contributes to tumor aggressiveness, therapy resistance, and reduced biomarker effectiveness[3]. Multi-regional sequencing has revealed the evolutionary processes underlying ITH[4,5], but the high cost and logistical constraints of these assays present significant barriers.

Even when multi-region sequencing is feasible (e.g., in large research studies), it is unclear how one should sample the tumor to efficiently capture ITH and study its evolutionary dynamics, especially in large tumors with high mutation rates where it is impractical to profile the entire tumor[6-8]. There is still much we do not understand about tumor evolution, highlighting a pressing need to explore alternative, scalable modalities for characterizing ITH.

An attractive alternative is to use tissue morphology from hematoxylin and eosin (H&E) stained histopathological slides that are already routinely used by pathologists to characterize tumors. Tissue morphology is used for diagnosis and for assessing tumor aggressiveness[9-11] and likely reflects phenotypic expression of molecular states in tumor cells. Under a microscope, extensive information about tumors can be appreciated, including tumor cell features, such as nucleolar prominence, variations in nuclear shape, and cytoplasmic characteristics. In addition, pathologists assess how tumor cells self-organize to form tumor architectures and how they relate to their microenvironment[12]. While the sheer scale and complexity of these features (~$10^5$ cells per slide) make quantification difficult[13], computational approaches offer an opportunity to systematically characterize ITH. Indeed, we and others have shown that applying deep learning (DL) approaches to histopathologic slides can deconvolute ITH by identifying driver mutational states, molecular subtypes, and diagnostically distinct areas[14,15]. However, our past work has focused on predefined aspects, rather than the totality of ITH.

**MAIN CONTRIBUTIONS**

Here, we introduce MorphoITH, a functionally unbiased framework to deconvolve ITH from histopathologic slides. We prototype this approach in clear cell renal cell carcinoma (ccRCC), a model for ITH[7,13]. Our approach is designed to overcome two key challenges. First, quantifying biologically meaningful morphologic differences and differentiating them from inconsequential fluctuations. We address this using a self-supervised DL model trained on the assumption that in a tissue microarray (TMA), areas in close physical proximity are more likely to be functionally similar. Second, mitigating biases introduced by spatial proximity: spatially proximal regions tend to be more similar (morphologically, molecularly and functionally), potentially introducing spurious statistical correlations even in the absence of a direct connection[16]. To address this, we incorporate statistical controls that explicitly account for the influence of spatial proximity, ensuring a more reliable evaluation of the relationship between morphology and molecular readouts.

We validated MorphoITH in several ways. First, we show its ability to capture biologically meaningful tissue types, vascular architectures, and nuclear grade changes. Then, we applied MorphoITH to evaluate the effect of key driver gene mutations in ccRCC (*BAP1/SETD2/PBRM1*), demonstrating that morphological heterogeneity reflects genetic heterogeneity and can guide efficient sampling strategies. Finally, using multi-regional sequencing data from three patients (26 samples total), we explored how tumor evolution drives morphological ITH, and observed that morphologic changes are typically incremental, although convergent evolution occasionally results in similar phenotypes across divergent genetic trajectories. These findings position MorphoITH as a scalable, unbiased tool for understanding the interplay between morphology and genetic heterogeneity, with implications for improving tumor sampling and advancing precision oncology.

# RESULTS

## MODEL DEVELOPMENT AND VALIDATION

Our goal is to develop a model that, given an input image patch, summarizes its morphological essence in a feature vector (Fig. 1A), which can be used for phenotypic deconvolution of tumor heterogeneity and the study of its biological underpinnings (Fig. 1B, C). However, as no two image patches are identical, we need the features to focus on the functionally relevant aspect of tissue morphology and ignore less consequential differences. For a general-purpose morphological descriptor, we cannot focus on specific functional classes or morphological traits but must learn the relevant phenotypes in an unbiased fashion. We reasoned that patches of nearby tissues in a TMA core are likely to be functionally similar and sought to perform self-supervised training such that patches from the same core had similar feature vectors, relative to those from different cores (Fig. 1A). In optimizing the model, we compared a variety of approaches commonly used in self-supervised training, including different: a) training frameworks requiring examples of positive pairs alone (BYOL[17]) or also negative pairs (Triplet[18], MoCov2[9]), ii) backbones (ViTb[19], ResNet50[20]), and iii) definition of positive pairs (Sup. Fig. 1A).

We benchmarked the models' ability to capture morphology on a held-out portion of the cohort (TMA training in Methods). First, we selected the best model based on a retrieval task: how often the nearest neighbors of a patch in feature space come from the same TMA core (Sup. Fig. 1B, C). Additionally, for reference, we considered two pretrained models: one for classifying natural images (ImageNet[21]) and the other trained on histopathological data in a self-supervised manner (RetCCL[22]). We noted that ImageNet trained on natural images achieved poor results, and the RetCCL approach trained on pan-cancer histopathology data outperformed it. However, the highest accuracy was achieved using kidney cancer patches with a BYOL training strategy and a Vision Transformer encoder. Thus, we decided to use it for the MorphoITH framework in all subsequent analyses. Next, we sought to evaluate how well MorphoITH performs as a more global similarity measure by comparing patches from different cores. Both in a nearest neighbor analysis and a global t-SNE representation, we found that when patches were considered similar but originated from different cores, they tended to have similar morphology, e.g., the same nuclear grade (Fig. 1A, Sup. Fig. 2A-C TMA validation in Methods). What is more, regions of heterogeneity within a single core (e.g., tumor vs. stromal patches) were accurately distinguished as distinct phenotypes (Sup. Fig. 2D). Taken together, these results suggest that our approach reliably captures morphologic differences.

## FRAMEWORK FOR QUANTIFYING MORPHOLOGIC DISTINGUISHABILITY BETWEEN REGIONS

Next, we extended this framework to analyze ccRCC whole slide images (WSIs). To quantify how well two regions differing in a predefined parameter (e.g., molecular state) are distinguishable based on their morphology, we developed a separability metric based on the performance of a naïve linear classifier distinguishing MorphoITH profiles (Fig. 1B; Separability Analysis in Methods). A key challenge in interpreting this separability is that physically proximal regions tend to be similar both morphologically and molecularly. To address this, we established two baseline distributions (Fig. 2A) for comparison: 1) a contiguity baseline that preserves the spatial structure of tissue regions, and 2) a true negative control where ground truth classes are randomly split to simulate intra-class morphological variability. Based on these baselines, we classified separability as: a) fully separable if it exceeded the 95th percentile of the contiguity control distribution, b) partially separable if it fell between the 95th percentiles of the contiguity and negative controls distributions, and c) not separable otherwise.

We first validated this framework by testing the ability of MorphoITH to distinguish tumor and non-tumor regions (black vs. grey areas in Fig. 2A: ground truth). We used a sliding window to generate 1000-dimensional morphological feature vectors across patches in a slide (Whole-slide Images Inference in Methods). To visualize spatial patterns of morphological similarity we used a pseudo-color visualization (Similarity Visualization in Methods) such that areas with similar MorphoITH features vectors are similar in color (Fig. 2A, last column). This approach distinguished tumor and non-tumor areas (Sup. Fig. 3A: tissue types form their own clusters, Sup. Fig. 3B: two separate clusters in the tumor region), suggesting that MorphoITH clustered morphologically similar regions. More quantitatively, on a cohort of 444 TCGA KIRC slides, our framework correctly separated tumor and non-tumor regions in 94% of cases (Fig. 2B, Sup. Fig. 4A). In 5% of the cases, separability was partial likely due to ambiguous tissue classification (e.g., regions consisting of tumor interspersed within stroma, Sup. Fig. 4B). The remaining 1% represent failures due to staining artifacts or tissue damage (Fig. 2B, Sup. Fig. 4C). Thus, MorphoITH was able to distinguish between normal and tumor areas within the same slide.

Next, we evaluated MorphoITH's ability to resolve clinically relevant morphologic features. We focused on tumor architecture and nuclear grade. We categorized recurring architectural patterns, identified based on spatial patterns of blood vessels, into 3 prognostic classes. Class 1: small and large nests, bleeding follicles; Class 2: alveolar, trabecular, papillary; and Class 3: solid [13,23,24]. At the cellular level, nuclear grade is a key clinical feature of ccRCC with significant prognostic implications. According to nuclear grade, ccRCC are classified into low (grade 1-2) and high (grade 3-4) tumors. Nuclear grading is based on tumor cell nuclear size and nucleolar prominence[25]. For our experiments, a pathologist in training (VP) manually partitioned tumor regions based on architecture and grade, with variations in both grade and architecture often seen within the same slide (Fig. 2C; WSI-1 and WSI-2 cohorts in Methods). MorphoITH fully separated areas of different architectural classes in 95% of cases, with small and large nests occasionally grouped together due to their morphological similarity[23] (Fig. 2D, Sup. Fig. 5A). Similarly, areas of low and high grade within the same slide were fully separable (Fig. 2E). However, the separation between grade 1 and 2 was partial in 75% (3/4) of cases (Sup. Fig. 5B), possibly because these are hard to distinguish as suggested by substantial interobserver variability[26,27].

Interestingly, MorphoITH was able to integrate multiple computational morphological descriptors (e.g., nuclear size, vascular density; Sup. Fig. 6, Morphological Descriptors in Methods) into its assessment. Essentially, areas considered similar by MorphoITH are also considered similar by all other measures, but the converse is not always true (changes detected by MorphoITH may not be detected by individual descriptors). This highlights MorphoITH ability to combine multiple morphological aspects into a unified similarity metric. In summary, MorphoITH offers a general-purpose measure of morphological similarity which captures a wide range of clinically relevant as well as other morphology features.

**MORPHOITH DETECTS SUBCLONAL CHANGES IN STATUS OF KEY DRIVER GENES**

Next, we explored the relationship between morphological heterogeneity, as assessed by MorphoITH, and genetic ITH. Rather than predict the status of a driver genes within histopathology slides[14], we compare areas within a slide differing in the driver gene status and ask if there are *any* differences in morphology, regardless of whether these are specific to that driver gene. For these experiments, we used matched slides with immunohistochemistry (IHC) results that accurately readout the state of three key ccRCC driver genes (*BAP1*, *SETD2*, and *PBRM1*) and H&E (WSI-3 cohort in Methods). We focused on slides with heterogeneity in these genes. Based on the corresponding IHC, a pathologist annotated "wild-type" (WT) and "loss" regions on H&E images. As expected, for each of these genes, we observed morphologic changes at the boundaries between wild-type and areas of loss (Fig. 3A). Next, we deployed MorphoITH more systematically across 34 slides with ITH in one of these genes. 58% showed full separability (distinct morphologies for WT vs. loss), 33% partial separability (sub-regions deviating from the broader genotype-associated morphology), and 8% no separability (Sup. Fig. 7A). Among the three genes we found the strongest morphological effects were seen in *BAP1*, and *SETD2* had better separability than *PBRM1* (Fig. 3B).

Interestingly, even within WT or loss regions, we often observed morphologically distinct sub-regions, suggesting local variation unaccounted by these driver genes. To investigate this further, we developed an automated approach that partitions slides into spatially constrained morphologic clusters using MorphoITH features (Spatially Constrained Clusters in Methods). Applying this approach to an annotated case with ITH in both *BAP1* and *SETD2*, we found that each gene loss region formed its own distinct cluster (Fig. 3C). Next, we reasoned that if indeed these regions represented "sub-clones", they would be completely contained within the WT or loss areas and not span across these areas. To test this, we first functionally defined individual clusters as being either within WT (>0.8 overlap with WT), within loss (<0.2 overlap with WT), or split across WT and loss (functionally defined as having between 0.2 and 0.8 of area overlap with WT; Fig. 3D; Overlap Analysis in Methods). Next, we tested how frequently our clusters were split across WT and loss, in comparison to randomly generated clusters based on our contiguity baselines. We found that for both *BAP1* and *SETD2* (*PBRM1* failed to achieve significance) there was a significant reduction in areas spanning WT and loss, consistent with the morphological clusters representing subclones (Fig. 3E, Sup. Fig. 7B).

To further understand how genetic ITH contributes to morphological ITH, we analyzed 1268 WSIs, comparing slides with both loss and WT regions for the three driver genes to those with homogenous gene status (Spatially Constrained Clusters in Methods). While most slides exhibited some morphological heterogeneity (Sup. Fig. 7C), possibly due to variation in other genes, those with ITH in *BAP1* and *SETD2* (*PBRM1* failed again to achieve significance) showed a higher degree of morphological heterogeneity than those with homogenous gene status (Sup. Fig. 7D). These findings suggest that *BAP1* and *SETD2* exert profound effects on morphology that goes beyond that from other factors.

Finally, following the idea that morphological distinct areas are in distinct molecular states, we wondered whether morphology guided sampling could help in more efficient capture of molecular ITH in a slide. Specifically, given a budget for number of samples we can capture, we propose a morphologically guided approach (Fig. 3F), where we partition the slides into the specified number of morphologically most distinct areas and capture one sample from each (Morphology Guided

Sampling in Method). We find that indeed, by sampling based on morphology, we can capture the genetic heterogeneity for *BAP1* and *SETD2* mutations much more efficiently than random sampling, for example more than doubling the likelihood of capturing both a WT and loss area with two samples.

**MORPHOLOGY SIMILARITY REFLECTS UNDERLYING EVOLUTIONARY TRAJECTORIES**

To better understand how tumor evolution, the driver of genetic ITH, is reflected in morphology, we analyzed multi-regional sequencing data from three patients. For these experiments, each patient had a set of matched whole-exome sequencing and H&E images for morphologically distinct areas within the tumor (average size=6.6mm$^2$; WSI-1 cohort in Methods). Unlike the analyses above, the areas were drawn from different slides to better represent the overall tumor heterogeneity (Fig. 1C). For each of these patients, we sought to assess the relationship between morphological differences (Fig. 4A-C t-SNE plots, Sup. Fig. 8-10) and their evolutionary relationships inferred from their phylogenetic trees (Fig. 4A-C trees based on single-nucleotide variants (SNVs) and arm-level copy number alterations; Phylogenetic Trees in Methods).

As the different samples were mostly morphologically easily distinguishable based on our separability analysis (Sup. Fig. 11), we compared morphology by visualizing patch features individually (Fig. 4 A-C, right column) and using the averaged patch features in an area (Fig. 4D, Sup. Fig. 9, right column, Sup. Fig. 12 and 13). Patient A displayed concordant evolutionary and morphological patterns with: a) an early divergence that led to a morphologically distinctive low-grade clone (T1a-A9), b) a branch containing two clones (T1e-A3, T1f-A3 with SNVs in *LRP1B*, *RNF213*) exhibiting similar morphologies, and c) a clade of 5 samples containing two metastases (M3a-A12, M3b-A4 with SNVs in *AXIN2* and *RGS7*) which separated out based on both genetic and morphological similarity. Patient B showed a comparable pattern with: a) an early divergence of a low grade region (T1b-A7) and metastatic branches (M1a-B5, M2a-B6 with *PBRM1* SNV), and b) a clade of 4 samples, with a sub-clade (T1d-A5, T1e-A5 with *PBRM1* SNV and 14q gain) branching further. Patient C, while largely reconciling genetic and morphological clustering, presented more genetic complexity among morphologically similar samples. For example, three groups: i) T1-A6, ii) T1b-A7/T1c-A7 and iii) T1f-A11/Th1-A5 were clearly separable both morphologically and genetically. However, among the remaining samples, which are all morphologically and genetically similar, T1e-A11 is a genetic outlier, which seemingly obtained this morphology via a highly distinctive early branch. Overall, these results demonstrate a strong alignment between genetic and morphological similarities across all three patients.

More globally, we find a strong correlation between genetic and MorphoITH similarity between areas within the same patient (Pearson correlation coefficient r=0.56, p=5.5E-10; Fig 4D). Notably, deviations from this trend provided key insights. Rare cases of morphologically similar samples with significant genetic differences were observed, exemplified by T1e-A11 in Patient C (Sup. Fig. 12). These events likely reflect convergent evolution. Conversely, substantial morphological differences with minimal genetic variance were often linked to highly impactful genetic alterations. For example, 11 outliers with high morphological differences but low genetic variance are pairs that all differ in *TSC1* status (Fig. 4D, Sup. Fig. 12). We see a similar trend, although with lower sensitivity, to conventional morphologic descriptors such as grade or architecture. While substantial genetic changes were typically needed for shifts in grade or architecture, regions with identical grades often exhibited significant genetic variability (Fig. 4E). We note that these trends are unlikely to be driven simply by slide-to-slide variations in staining: we consistently observe samples from the same slide being separated while those from different slides being grouped together in ways that are consistent with grade and architecture (Sup. Fig. 10). More importantly, staining normalization does not influence our results (Sup. Fig. 13). Taken together, a parsimonious explanation of these observed patterns is that newer clones emerging during tumor evolution typically show incremental changes in morphology, although occasionally distinct genetic trajectories can converge on the same morphological phenotypes. This nuanced relationship between genetic and morphological heterogeneity underscores the utility of MorphoITH in characterizing tumor evolution and phenotypic diversity.

## **DISCUSSION**

Tumor evolution is arguably the fundamental challenge in oncology. The emergence of subclones in different genetic and epigenetic states makes it challenging to both diagnose and treat cancer. Recent multi-region sequencing studies have underscored the inadequacy of profiling single tumor regions in understanding the complexity of most solid tumors[8,28]. However, the cost associated with multi-region sampling limits their scalability. As tissue morphology reflects the underlying molecular state, we hypothesize that morphological variation in routinely used histopathologic slides, if quantifiable, could serve as a cost-effective proxy for molecular/functional heterogeneity.

Here, we present MorphoITH, a deep learning framework designed to: a) identify areas of similar and distinct morphology, and b) assess whether these differences relate to spatial variation in other biological properties. We developed MorphoITH in the context of ccRCC, which is a paradigm for intra-tumor heterogeneity[4]. Areas identified as similar by MorphoITH were related; they came from the same TMA core, shared nuclear grade and vascular architecture, or shared other computational image-derived features. By applying MorphoITH to WSIs, we observed abrupt morphological transitions that often correlated with changes in driver mutation status, such as *BAP1* and *SETD2* loss. We also demonstrated how these findings could guide tumor sampling to capture genetic heterogeneity. Additionally, our analysis of multi-region sequencing in three patients suggests that genetic divergence often leads to morphological changes. Furthermore, evolutionary trees derived from genetic data aligned well with morphological evolutionary trees. Interestingly, particular genetic events were associated with abrupt morphologic changes. In addition, we observed convergence on similar morphologies from distinct evolutionary trajectories. These findings underscore the complementarity of morphological and genetic analyses to dissect tumor progression.

A key differentiator of MorphoITH is its unbiased nature. While previous studies have used morphology of histopathologic slides to characterize ITH, they have largely focused on specific morphologic phenotypes (e.g., grade in ccRCC[29]). Some such studies have sought to predict a predefined disease state (molecular subtypes in breast cancer[30] or assess heterogeneity[31]). In contrast, MorphoITH aims to comprehensively capture morphological changes, serving as a general-purpose framework to explore ITH. To enable MorphoITH to assess morphology without relying on predefined labels, we trained a self-supervised model with hyperparameters[22,32-35] optimized for this task.

To validate its biological relevance, we established synthetic baselines to distinguish meaningful correlations from trivial spatial autocorrelations (arising purely from spatially proximal areas being similar[16]), adapting strategies from single-cell and multiplex imaging studies[36,37]. Using these, we were able to assess how morphology served as a proxy for biologically relevant (genetically informed) ITH.

A particularly striking finding was the reliable assessment of abrupt transitions in contiguous areas of ccRCC. Abrupt transitions characterize RCC, in contrasts with other cancer types, where the changes are more gradual and thought to be epigenetically driven[38]. Our data suggest that these abrupt transitions represent genetic "clonal" evolution. Specific examples include the focal loss of *BAP1* and *SETD2* (and to a lesser extent *PBRM1*). When abrupt transitions did not align with *BAP1/SETD2* changes, they typically behaved like "subclones" in being confined solely to the loss or WT regions.

We note the distinction between using MorphoITH, which tests whether a subclonal genetic event (e.g., *BAP1* loss) produces *any* morphological change in a given slide, to our previous approach, which predicted which regions in a slide exhibited *BAP1* loss by identifying morphological signatures specifically associated with this event across all slides[14]. In other words, MorphoITH tests the strength rather than uniqueness of the impact of a genetic event on morphology. Multiple genetic events, such as those acting on different levels of the same pathway, may impact morphology similarly and would all be picked up by MorphoITH, but fail to be distinguished by a classifier. This may explain why, in contrast to our previous work, we see a stronger link with morphology for *SETD2* which is associated with higher grade tumors than *PBRM1* (whose deficiency is not associated with an increase in grade or decrease in vasculature)[11,39]. Based on this idea that MorphoITH can identify areas that are likely to be genetically distinct, we used it to develop a morphology guided sampling scheme for detecting driver gene heterogeneity through multi-region sequencing, which otherwise either requires on average 7 randomly sampled areas to capture 75% of driver gene mutations[7,8].

Finally, our multi-regional sequencing dataset allowed for preliminary investigation into how tumor evolution is reflected in morphologic heterogeneity. In ccRCC, tumor grade, provides only a coarse, monotonic measure of tumor progression. We[13] and others[23] have hypothesized stereotypical trajectories between well-defined vascular architectures, and similar trends have been identified in other cancers[40]. Rather than focus on these predefined states, here, we investigated whether the morphology similarity between regions from the same patient could provide insights into their evolutionary relationships. Our findings suggest that tumor evolution frequently manifests as incremental morphological changes that reflect genetic divergence. However, we also found clear exceptions to this alignment. For example, potent genetic events like *TSC1* loss, which is typically associated with mTORC1 activation, were associated with abrupt and drastic changes in morphology. As such, our preliminary findings reveal compelling links between tumor evolution and morphology, highlighting the potential of integrating morphological and genetic analyses to better understand tumor heterogeneity and progression.

There are many avenues to expand upon this work. Firstly, larger studies, such as those leveraging multi-region sequencing datasets like TRACERx Renal[4,7], are needed to validate our observations. Secondly, future work could explore the relationship between morphology and a broader range of molecular pathways by leveraging spatial transcriptomics. Thirdly, emerging evidence suggests that cellular morphology can actively influence signaling events expanding upon the concept of structure conforming to function[41]. Thus, MorphoITH may set a foundation for deeper biological insights. Finally, MorphoITH's could be expanded to other cancer types by utilizing foundation models trained in a pan-cancer context as encoders and adopting techniques to mitigate potential slide-to-slide staining variations[42].

In summary, MorphoITH provides an unbiased, scalable, and versatile framework for studying tumor heterogeneity. Its ability to capture morphology and genetic heterogeneity has practical implications for improving sampling strategies and understanding tumor evolution, offering a new lens to unravel the complexity of cancer biology.

## MATERIALS AND METHODS

### DATASETS

We make use of the following datasets in the paper. Each consists of H&E-stained slides from Formalin Fixed Paraffin Embedded (FFPE) tissue blocks. The slides were scanned at either 20X or 40X resolution, and as all analyses were performed at 20X, images were downsampled as needed.

1. **TMA training:** the full training dataset has 10 tissue micro array blocks, 785 cores total, from ccRCC patients surgically treated at UTSW and representing all stages of the disease[43,44].
2. **TMA validation:** the retrieval task used for comparison of MorphoITH with RetCCL and ImageNet was done on a separate ccRCC TMA dataset that included patients treated for high stage ccRCC at UTSW[45]. The dataset consisted of 3 blocks, 238 cores total, that had their core-level nuclear grades assigned by pathologists.
3. **WSI-1**: a UTSW cohort of 41 ccRCC WSIs (UTSeq Data[46]) was used to: 1) annotate by pathologists at least two different nuclear grades sites (16 WSIs) and at least two different vascular architectures sites (26 WSIs) per slide, b) perform multi-region sequencing on (18 WSIs). The sampling sites for multi-region sequencing were annotated on the WSIs. Number of samples per patient: Patient A=8 (6 primary sites, 2 small intestine metastatic site), Patient B=7 (5 primary sites, 2 pancreas metastatic site), Patient C=11 (8 primary sites, 2 adrenal metastatic sites (M2), 1 lymph node metastatic site (M1)).
4. **WSI-2**: 10 ccRCC WSIs (UTSW CD31 Re-stain[46]) was used to: 1) annotate by pathologists at least two different nuclear grades sites (4 WSIs) and at least two different vascular architectures sites (7 WSIs) per slide, 2) analyze handcrafted features from tumor areas of whole slide images (10 WSIs).
5. **WSI-3**: 1268 ccRCC WSIs from Mayo Clinic[47] that had known driver gene status of wild-type (WT) or loss, as assessed by immunochemistry (IHC) assays done on serial sections, were utilized as an input for heterogeneity score calculations. Subset of cases that exhibited areas with both loss and WT driver gene status within the same slide (18 *BAP1*, 9 *SETD2*, and 9 *PBRM1*) were additionally annotated for their localized (focal) loss.
6. **TCGA**: After image quality control as described previously[46] and additional filtering of slides with insufficient tumor area (Whole-Slide Inference in Methods), 444 WSI ccRCC slides from the Pan Cancer Atlas study[48] were used to assess separation of tumor from non-tumor.

| Dataset | Number of slides |
|---|---|
| TMA training | 10 (785 cores) |
| TMA validation | 3 (238 cores) |
| WSI-1 | 41 |
| WSI-2 | 10 |
| WSI-3 | 1268 |
| TCGA | 444 |

**TRAINING DATA**

The models were trained on patches that were 224 x 224 pixels at resolution 20X (0.5 microns per pixel). The training dataset (TMA training) yielded a total of 785 cores and 153,968 patches, extracted by random selection of maximum 200 patches per core with overlap allowed. During the retrieval task used to compare different models' architectures and backbones in order to choose the most suitable ones, the models were trained on 2/3 of the training dataset and validated on the remaining 1/3. The training task involved feeding the models pairs of patches that were either an augmented pair of: the same patch (input type: same), two different patches from the same TMA core (input type: close), or a combination of the aforementioned pairs (input type: combined). If the model required negative examples, those were defined as patches from other TMA cores. The final encoder used for MorphoITH was trained on all the available patches, with the combined input type.

**MODEL SELECTION AND COMPARISON**

We compared different deep learning architectures, encoders, and input types. For architectures, we tested: 1) BYOL[17,49,50] that did not need any negative examples (patches considered to be different) for training, as well as 2) triplet (with TripletMarginLoss[18]) and 3) MoCoV2[51] that required negative examples. In the case of BYOL and Triplet, their backbones were either Vision Transformer (ViTb[19,52]) or ResNet50[20] pretrained on ImageNet[53]. Training was done with batch size 32 and Adam optimizer with a learning rate of 3e-4. Output of the last linear layer of the models was 1,000-dimensional. Input patches were augmented using HED and color jitter, random gaussian blur and elastic, and random rotation[54]. Additionally, color saturation was reduced with random multiplier.

Final model selection and comparison with available off-shelf models was performed based on results from a retrieval task, in which we checked how often the nearest neighbors of a patch came from the same TMA core based on cosine distance between feature vectors. Based on these comparisons, the final MorphoITH encoder was selected to have BYOL architecture and ViTb backbone. Additionally, as controls, RetCCL with weights as provided by the authors[22] and a frozen Resnet50 pretrained on ImageNet were used for profiles generation.

**WHOLE-SLIDE IMAGES INFERENCE**

Inference was performed on 224px patches tessellated across the slide in steps of 100px to generate a grid of profiled patches for: 1) MorphoITH, which generated 1,000-dimensional feature vectors for each patch, and 2) a pretrained in-house region classifier, which predicts patches as tumor, normal, stroma, blood, necrosis, immune, and background. Tumor areas with less than 100 tessellated patches were discarded. If the regions model's output was used to create ground truth for region classification, its activation outputs were smoothed using a gaussian filter with size 3.

We first used the region classifier as ground truth to test MorphoITH's ability to distinguish tumor vs non-tumor regions. For all subsequent tasks, we focus purely on the MorphoITH profiles from tumor regions (non-tumor filtering) while also discarding areas with pen-marks identified using an in-house artifact detection model.

**SEPARABILITY ANALYSIS**

We first obtained ground truth assignment of areas on WSI images into different classes. These assignments were based either on pathologists' annotations of the areas on WSIs (for architecture, grade, and driver mutation status which was manually transferred over from a corresponding IHC slide), or on the output of an in-house tissue classifier. Given such ground truth assignment of areas from two classes (e.g., tumor vs not), we created a measure to assess their separability based on MorphoITH features.

First, the 1,000-dimensional MorphoITH outputs were reduced to 10-dimensions using PCA. Then, a linear support vector machine (SVM) classifier with class balancing was trained using 50% of the patches to distinguish the two classes, and the accuracy of the model on the held out 50% was used as the separability measure. In cases where there was more than one class present in a slide, we iteratively focused on each pair of classes.

In order to account for spatial autocorrelation, we generated two sets of control ground truth classes by: a) contiguity baseline: randomly creating circular annotations within the regions, while maintaining the area size of each class, and b) negative control: splitting each contiguous region for a class in half and giving each half different label. We generated 10 random control-ground truth per each annotated class pair in a WSI, calculated their separability scores, and pooled across all WSIs. We then calculated the 95$^{th}$ percentile for each of these distributions.

A (non-control) separability scores was assessed as having: a) full separability: if it was above 95$^{th}$ percentile of contiguity baseline, b) partial separability: if it was between 95$^{th}$ percentile of the two controls, and c) no separability if it was below the 95$^{th}$ percentile of negative control.

**SIMILARITY VISUALIZATION**

As 1000-dimensional outputs of MorphoITH on WSI are difficult to interpret, we developed a visualization which assigned patches in a WSI with similar MorphoITH features to similar colors. To achieve this, we performed a PCA of the MorphoITH features, and for each patch took the 3 first principal component loadings, scaled them to a [0,1] range and use them to define colors in a RGB color scheme.

**MORPHOLOGICAL DESCRIPTORS**

MorphoITH's similarity measure was compared to the following morphological descriptors: 1) nuclear sizes calculated from StarDist nuclear detection[55,56], 2) vasculature density from an in-house vascular model[46] that outputs vascular masks, 3) eosin intensity for non-nuclear pixels obtained by performing RGB to HED color space conversion guided by the previously mentioned nuclear mask. For a chosen pair of patches, their similarity in the three descriptors was calculated by taking their absolute difference, whereas MorphoITH's similarity was based on a cosine similarity between its 1,000-dimensional feature profiles. For each slide, we chose 100 pairs of patches based on their similarity in the aforementioned measures. To exclude patches being similar because they share some cells, we required that pairs of patches to be apart in distance by at least 10 other patches in a WSI. Those pairs were then compared to the similarity distributions of 100 randomly selected pairs, in order to get rank of their similarity.

**SPATIALLY CONSTRAINED CLUSTERS**

To obtain morphological clusters based on MorphoITH-derived similarity measure, we performed agglomerative hierarchical clustering on feature vectors, with: 1) set number of clusters N=10, 2) "ward" linkage, and 3) connectivity matrix to ensure spatial constraint of the clusters, as for our analyses we require a cluster to maintain continuity across a slide. A representative feature vector for each cluster was calculated by finding a feature vector that was closest in cosine distance to a mean cluster representation. The heterogeneity score between clusters was defined as cosine distance between the clusters' mean feature vectors, with clusters having areas below the 5th percentile or above the 95th percentile excluded to minimize the influence of potential outliers. To define a slide's heterogeneity score, we chose the highest score from those calculated between all its clusters.

**OVERLAP ANALYSIS**

We devised a measure to quantify the agreement between the following two partitions of the tumor region of a WSI: 1) loss/WT for *BAP1/PBRM1/SETD2* as described in the WSI-3 description above, and 2) 10 spatially constrained morphologic clusters. Specifically, for each cluster we assessed what proportion of its area was in the loss area. To account for registration errors, in performing this calculation we discarded the area that was labelled at the WT/loss boundary (obtained after dilation of the boundary between the annotations by kernel (5,5)). Then, we considered the proportion of a cluster within the loss region, and within the wild-type region. We considered a cluster to be split between the WT/loss region if its overlap was in range of 0.2-0.8. Finally, we calculated the proportion of clusters that are labeled as split using ground truth annotations, and those labeled as such when using control-ground truths. Controls were repeated for each slide 10 times. To compare whether the proportion of our splits is different (lower) than in case of the controls, we used Fisher's exact test.

**MORPHOLOGY GUIDED SAMPLING**

Our goal was to use morphological heterogeneity detected by MorphoITH to propose areas within a slide to profile for multi-region sequencing to better capture genetic heterogeneity. We tested this approach on slides from WSI-3, with intra-slide genetic ITH in *BAP1/PBRM1/SETD2*, where the goal was to extract punches with loss and WT with a minimum set of areas profiles. Given a budget of N areas to profile, our recipe is to identify the N most morphologically distinct regions based on MorphoITH, and profile representative areas from each. This was achieved by performing Spatially Constrained Clustering (as described above) with N clusters, and choosing a representative patch for each. If for the N chosen representative profiles their corresponding labels included both loss and wild-type, we considered the heterogeneity capture a success, and quantified the overall performance across slides in terms of the fraction of successes. Similar to the Overlap Analysis, we discarded areas labeled as "at boundary". Random sampling was performed by randomly choosing a representative patch from the tumor area. To test whether MorphoITH improved the success rate over random sampling for the same number of punches, we used the Mann-Whitney U test.

**SOMATIC VARIANT CALLING FROM WHOLE-EXOME SEQUENCING**

WSI-1 cohort has whole exome sequencing (WES) of FFPE tissues from 3 ccRCC patients each represented by multiple tumor regions and one adjacent normal site. Raw reads FASTQ data was processed using the SCHOOL pipeline[57] to trim reads, align to human reference genome GRCh38 (hg38), and call somatic variations by paired tumor and adjacent normal samples. The pipeline used stringent criteria to only output high confidence variations, requiring single-nucleotide variations and small indels to be reported by three callers: Mutect2 (gatk version 4.1.4.0)[58], freebayes (version v1.2.0)[59] and Strelka2 (version 2.9.10)[60]. Somatic allelic copy number variations (CNV) were called on the same paired tumor and normal WES samples using the FACETS (version 0.6.2) and FACETS-suite (version 2.0.8)[61] R packages. Chromosome arm level gain or loss was called when >50% of the chromosome arm had copy number gain or loss.

**PHYLOGENETIC TREES**

Both somatic synonymous and non-synonymous variations as well as chromosome arm-level copy number variations were used to construct phylogenetic trees. For each patient, we identified the entire set of observed variant events across all samples and used it to construct a variation matrix (variant x sample). In this matrix, events associated with mutations were assigned as 0 for wild-type and 1 for mutants (VAF>0). For arm-level CNVs, events were specified as -1/0/1 (loss/neutral/gain). The column corresponding to the normal sample had all 0s. Pairwise hamming distances between samples based on this matrix were used as our "Genetic Distance" and used to construct a phylogenetic tree based on the nearest-neighbor algorithm, which was further refined by nearest-neighbor interchange (NNI) rearrangement optimized for maximum parsimony. Finally, the tree was re-rooted to start from the normal sample. Trees were visualized in Fig. 4 in ultra-metric format (tips of the tree are equidistant from the root), while in Sup. Fig. 8 the trees were visualized as phylograms (tree lengths are proportional to the amount of change). Order of the parallel branch splits was rotated manually (while respecting the phylogenetic relationships in the tree) to better match the corresponding t-SNEs. Alterations likely to be important (COSMIC Cancer Gene Census curated list of oncogenes[62] and TRACERx Renal signature CNVs[7]) were marked on the corresponding tree branches. We only displayed alterations next to a branch if they were: a) in coding regions, b) found in all samples along that branch, and c) absent in samples diverging in other directions.

**MORPHOLOGICAL SIMILARITY FOR EVOLUTION**

MorphoITH features for all samples per patient were extracted as 224 x 224 pixels patches, with maximum of 250 patches per sample, either without normalization (Fig. 4), or normalized (Sup. Fig. 13) using Macenko[63] or Vahadane[64] schemes to match the color distribution of the slides from a patient. The features were plotted as t-SNEs with perplexity equal to 500 to visualize the trajectories of morphological evolution. For further analyses, mean representations per sample were calculated and compared by measuring cosine distances between pairs of samples ("MorphoITH distance").

**CODE AND DATA AVAILABILITY**

The source code will be available at the Rajaram Lab's public GitHub page upon publication. H&E images for TCGA KIRC can be downloaded from the TCGA GDC portal. TMA validation and TMA training can be accessed under TMA1 and TMA2, respectively, at https://doi.org/10.25452/figshare.plus.19324118. WSI-3 can be accessed at

https://doi.org/10.25452/figshare.plus.19310870. WSI-1 and WSI-2 cohorts are available from authors upon reasonable request.

**ACKNOWLEDGEMENTS**

We thank the patients who donated their tissues for analyses. A.W.N. is supported by The Cancer Prevention and Research Institute of Texas (CPRIT; RP210041). S.R. is supported by CPRIT (RP220294). J.B. and P.K. are supported by NIH (Specialized Program in Research Excellence in Kidney Cancer P50 CA196516).

**Figure 1.** Overview of MorphoITH, a framework for characterizing morphological similarity from histopathological slides. A) MorphoITH uses a self-supervised deep learning model to measure similarity learnt by taking input patches from the same tissue microarray (TMA) cores and training the model to produce similar feature vectors. Despite being self-supervised, the resulting features are biologically meaningful, e.g., image patches (points in scatter plot) of similar nuclear grade (colors) tend to have similar MorphoITH profiles. B) It incorporates statistical analyses to assess how well morphological heterogeneity (top row: false color depiction of MorphoITH profiles) relates to heterogeneity in a given parameter of interest (bottom row: in black). By comparing the separation to randomized baselines that were developed to account for the spatial auto-correlations inherent to tissue, we can categorize the separation as fully separable (Full), partially separable (Partial), and not separable (None). C) The developed measure of morphological similarity was additionally applied across samples within the same patient in multi-regional sequencing dataset to assess the relationship between evolution and morphology.

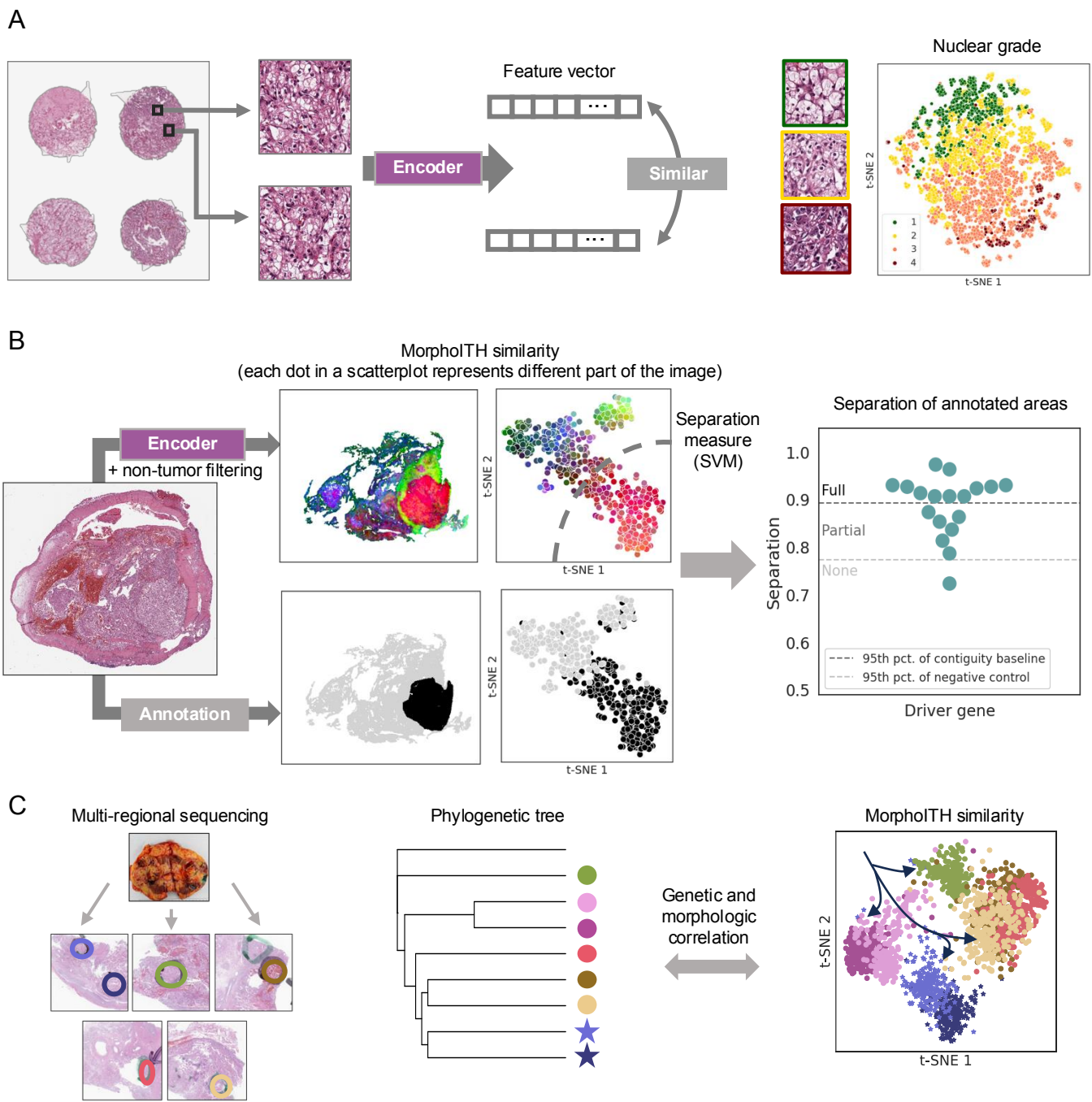

**Figure 2.** MorphoITH recognizes variance in tissue types, vascular architectures and nuclear grade in ccRCC. A) Given ground truth annotation of property of interest (top left), we calculated its separation scores (bottom left; SVM accuracy) in MorphoITH feature space (pseudo color on right, and t-SNE on bottom row). To control for separability arising solely from regions being physically distant, we established two baselines. The contiguity baseline, built by randomly adding circles with areas corresponding to those of ground truth annotations, validates our approach against random contiguous annotations. Negative control ensures that variance within annotations is comparable to that across them. Separability scores are constructed for multiple random locations of these baselines, and exceeding the 95th percentile of the contiguity and negative baselines defines Full/Partial and no overlap. B) Separation (y-axis) of tumor vs non-tumor tissue across whole slide images (individual points) from TCGA KIRC. Points were assigned as Full/Partial/No overlap with tumor region masks (as explained in A). Two examples of tumor areas of samples that were not separable with image artifacts (tearing, blurring) are shown. C) An example of a whole slide image with areas differing in nuclear grade and vascular architecture. Example images of the slide's architectures are visualized with frame colors corresponding to the legend. We plot a t-SNE with 600 randomly sampled patches, which shows how MorphoITH-derived feature space separates both grades and architectures. D, E) Separation of: D) vascular architecture, E) nuclear grade, across multiple slides as compared to contiguity baseline and negative control. Heatmaps show average separation of the ground truth annotations (upper half) and average contiguity baseline control (lower half). Strip-plots include comparisons of separation (y-axis) between pairs of annotations, now with coarser labels (x-axes): architecture class 1 (small and large nests, bleeding follicles) vs class 2 (alveolar, trabecular) vs class 3 (solid), and low grade (1-2) vs high grade (2-3). Each point represents a comparison between a pair of annotations within a slide.

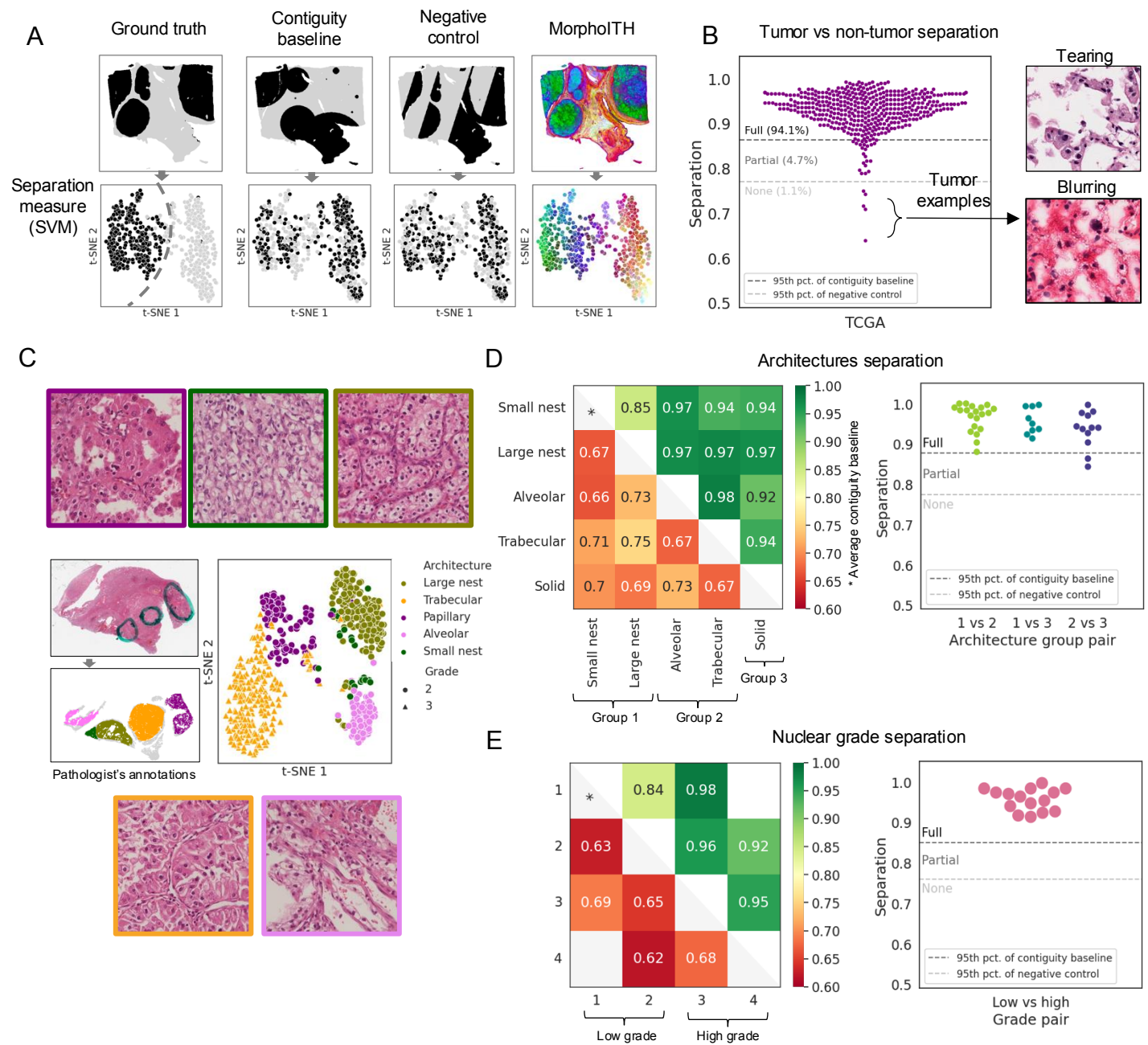

**Figure 3.** MorphoITH detects morphological changes in regions with loss in driver gene function. A) Example of tumor areas that exhibit distinct areas with driver gene in wild type and loss functional states. We show three driver genes (columns) and their ground truth based on annotations from custom IHC assays, H&E images, and MorphoITH output visualization using pseudo-colors that indicate morphological similarity (rows). B) Separation measurements across multiple slides with different driver gene statuses (each dot represents separation for one slide). 58% of cases have a full separation, 33% partial separation, and 8% no separation. C) The relationship between MophoITH based spatially constrained morphological clusters and genotype in a case with heterogeneity in both *BAP1* and *SETD2*. Examples images of morphologies of *BAP1* and *SETD2* loss are shown with frame color corresponding to the legend. D) An example of a H&E WSI (1st column), pseudo-color visualization of MorphoITH in its tumor regions (2nd column), spatially constrained morphological clusters (3rd column, N=10). The morphological clusters are then checked for overlap (4th column) with loss region in ground truth (5th column), which indicates areas of wild type and loss of driver mutation. E) Percentage of morphological clusters spanning both WT and loss regions (defined as overlap <0.2 or > 0.8 with loss region) for different driver genes. We compared the overlap of annotated loss areas (ground truth, purple) with corresponding controls (grey). F) Morphology guided sequencing. Bottom: given the number of sampling site N, we perform spatially constrained morphologic clustering with N clusters and select the most representative (Spatially Constrained Clusters in Methods) patch as a sampling site for the cluster. Top: Probability of capturing both (loss and WT) driver gene states for *BAP1* (p=4.4e-8), *SETD2* (p=1.7e-3), *PBRM1* (p=0.46), across all slides (mean plotted with the 95% confidence interval margins). We compared the sampling results to those captured by randomly choosing points within tumor areas (random sampling, grey).

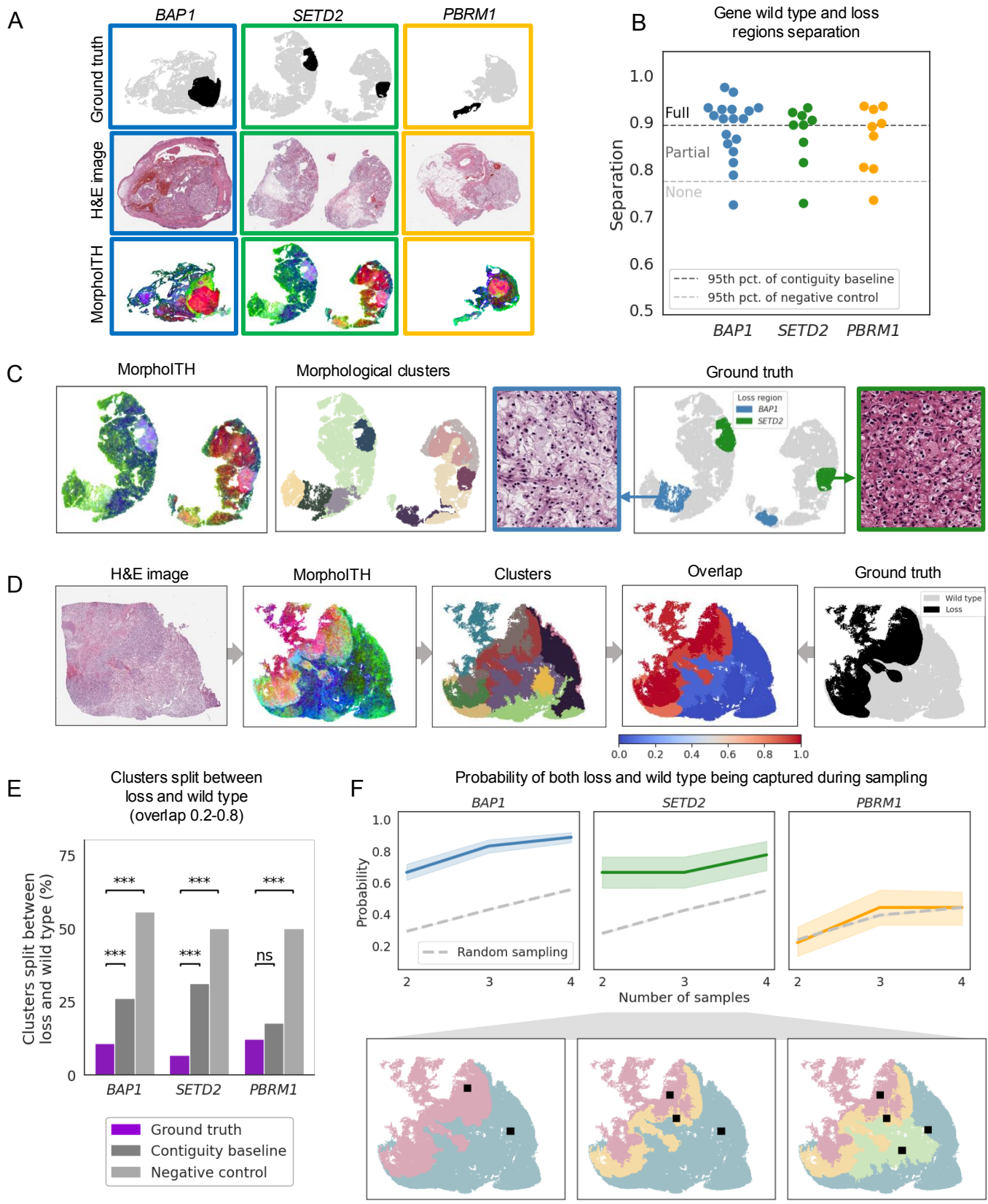

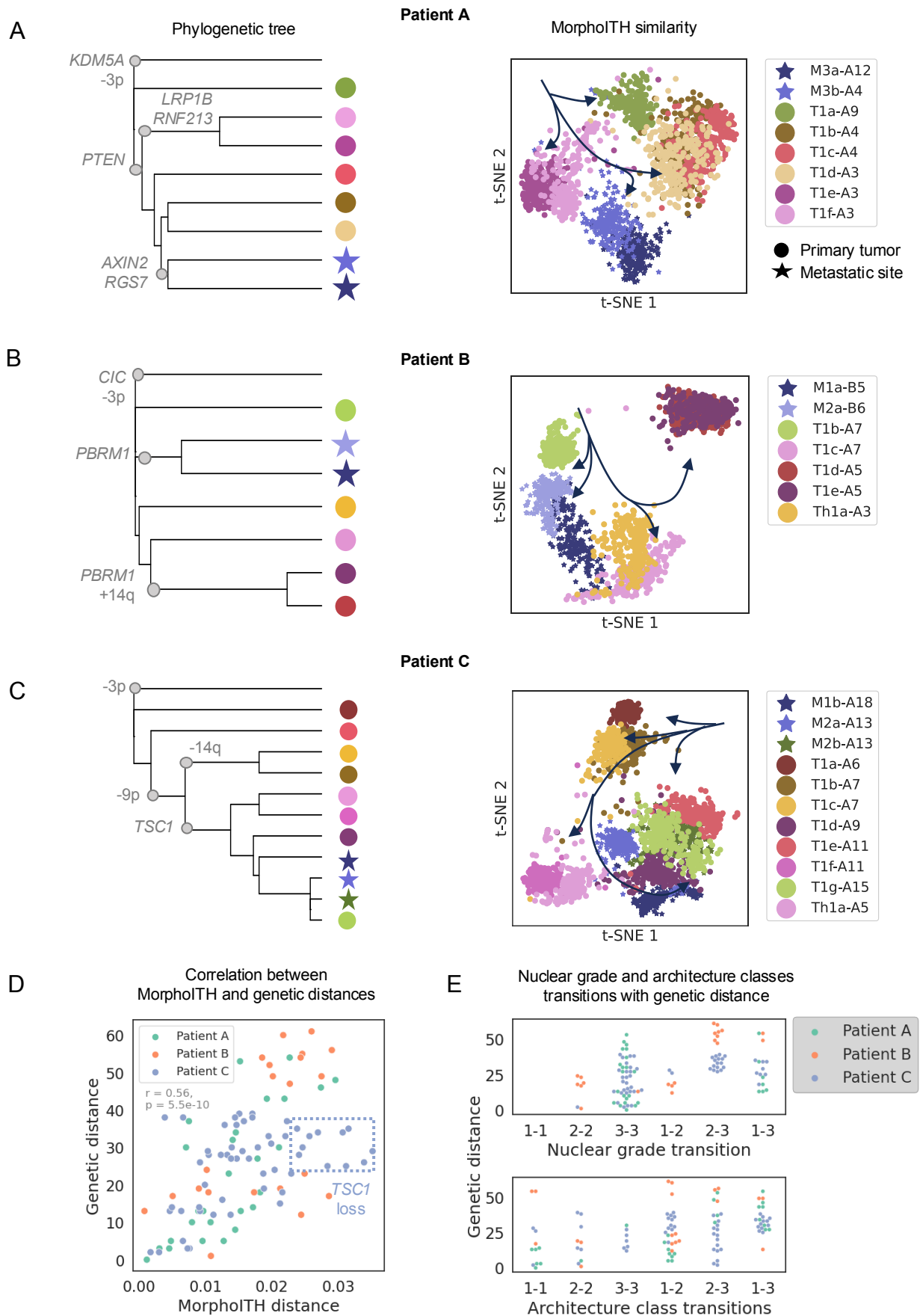

**Figure 4.** Morphology as a tool to deconvolve genetic heterogeneity. A-C) We compared phylogenetic trees (Phylogenetic Trees in Methods) for three patients (1st column) with their corresponding morphological similarity (visualized using t-SNE in 2nd column) based on the matched sequencing sites and H&E images. Each point in a t-SNE plot is a patch with point color corresponding to its sample site. Primary tumor and metastatic sites are indicated by circles and stars, respectively. In the t-SNE, the samples are named as: T/Mx-block number, with T = primary tumor, and M = metastatic. D) Global correlation between genetic distance (y-axis) and MorphoITH distance (x-axis) for all patients. Pearson correlation coefficient r = 0.56, with p-value = 5.5E-10. Each point is a pair of samples within a patient that are being compared. E) Distribution of genetic distances between samples as a function of clinically used morphologic measures. We check how the genetic distance (y-axis) changes when the samples do or do not differ in grade (upper), or in architecture class (lower). Same grade/architecture class transitions are denoted as 1-1, 2-2 and 3-3, while transitions across grade/architecture class include grades/classes 1-2, 2-3, 1-3 (x-axis).

**Supplementary figure 1.** Comparison of training strategies for MorphoITH development. A) Description of different approaches that were compared using the retrieval task. The approaches differ in model backbones (ViT, ResNet50), training strategies (BYOL, Triplet, MoCo-v2, pre-trained), and input types ("combined", "same", "close", and pretrained ImageNet or RetCCL). B) Comparison of the retrieval task, in which we calculate the fraction of patches (y-axis) for which the N-th nearest neighbor (x-axis) were from the same TMA core for different strategies. C) Retrieval task results (y-axis, fraction of patches captured from the same TMA core) at the 10th nearest neighbor for different strategies (x-axis).

A

| Backbones | Training strategy | Input type (positive pair definition) |
|---|---|---|
| ViT | BYOL | Combined* |
| ResNet50 | Triplet | Close* |
| | MoCoV2 | Same* |
| | Pre-trained | ImageNet (pre-trained) |
| | | RetCCL (pre-trained) |
| * Training Data in Methods<br>Same = same patch augmented<br>Close = close spatially patches augmented<br>Combined = combination of the above | | |

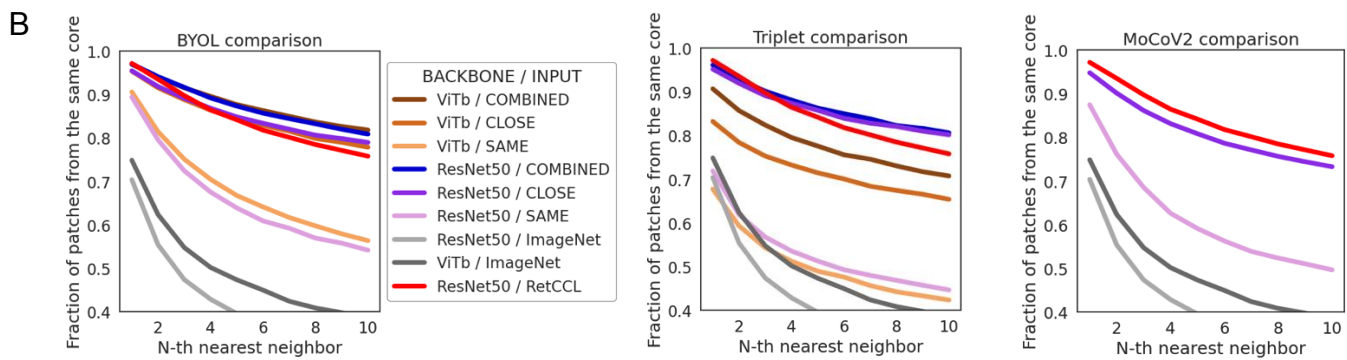

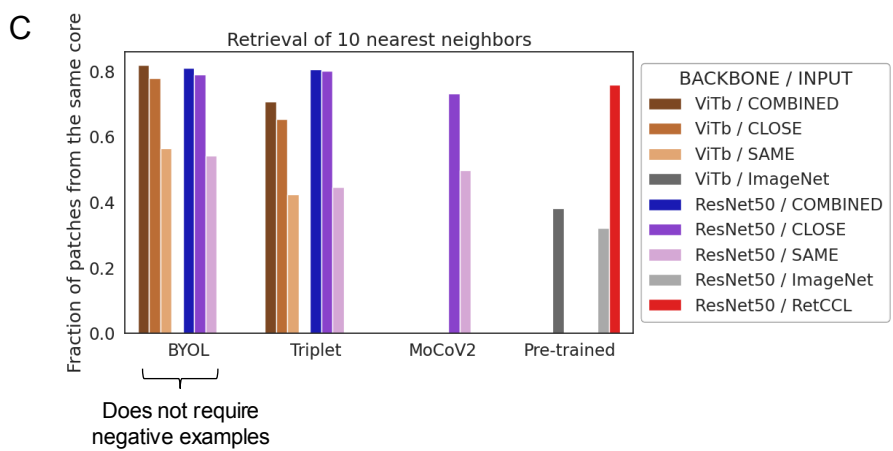

**Supplementary figure 2.** MorphoITH-based similarity captures biologically relevant aspects of morphology. For each patch in the independent "TMA validation" cohort, we ranked all other patches based on their similarity. A) An example of one patch and its 9 closest neighbors (G=grade, N=nearest neighbor rank). B-C) MorphoITH similarity across TMA cores and grades. B) We plotted the fraction of patches (y-axis, stacked fractions adding to 1) for which the N-th nearest neighbor (x-axis) were from the same TMA core (orange), have the same nuclear grade (but from different core, blue) or differ in both (grey). C) Box-plot showing the average nearest-neighbor ranks of patches from each of these classes. D) Global view of MorphoITH's feature space: t-SNE based dimensional reduction was used to visualize the similarity between the MorphoITH vectors for a random subset of 10,000 patches. Each point represents a patch from a TMA core. Points are colored based on the core they were derived from, with patches from one example core highlighted in red and the corresponding patch image depicted on the side. Points' positions are the same as in Fig. 1A, except there patches are colored by grade.

A

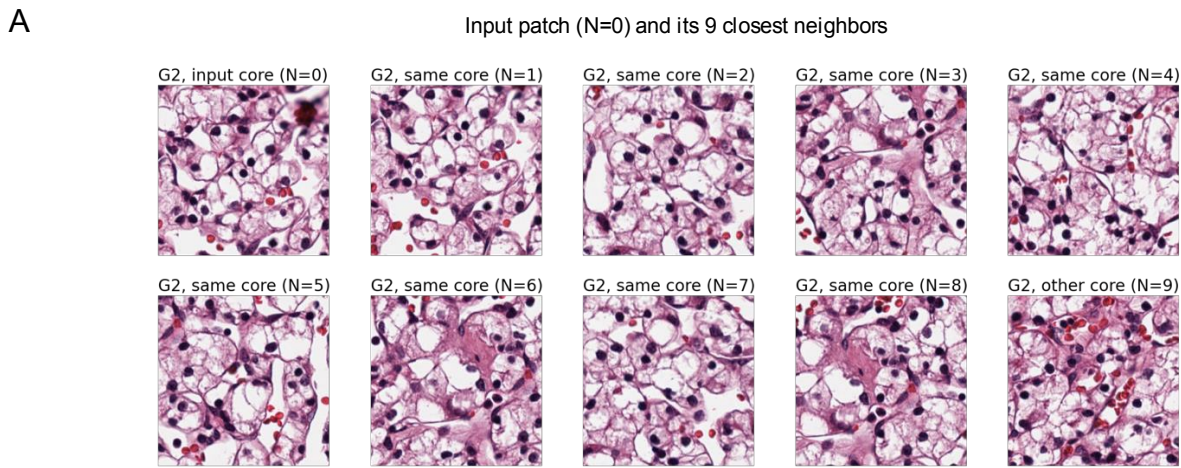

Input patch (N=0) and its 9 closest neighbors

B

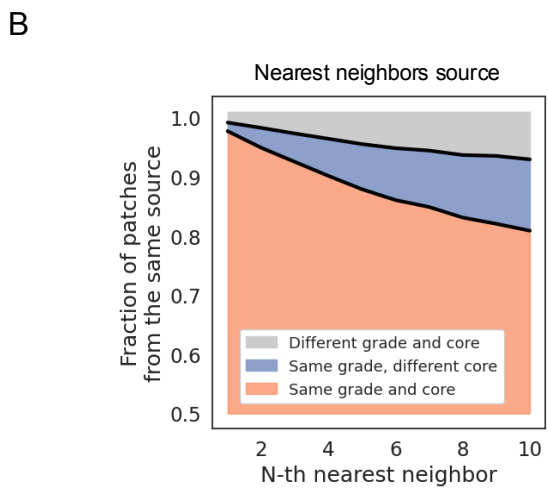

C

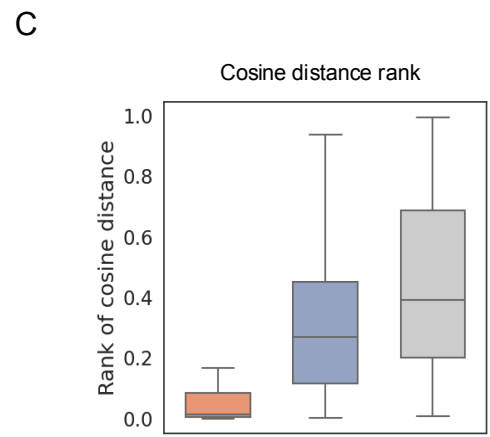

D

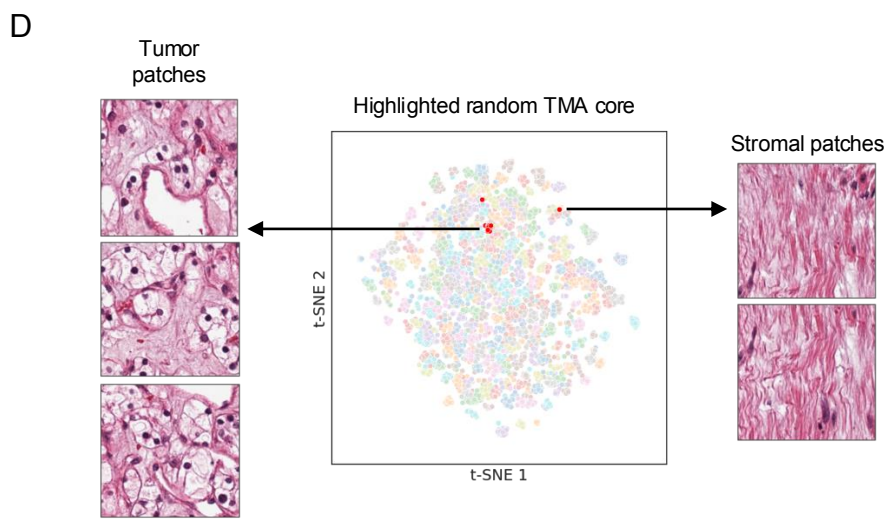

**Supplementary figure 3.** MorphoITH visualization of tissue-type deconvolution. A-B) Two examples comparing whole slide MorphoITH profiles to different tissue region types. Tissue type ground truth (1st rows) was obtained using an in-house deep learning "Region Classifier". MorphoITH features were extracted across patches in the slide (2nd rows) and visualized with pseudo-colors, where similar colors indicate similarity in morphology. Separation of tumor vs non-tumor areas was performed on MorphoITH-derived feature vectors using an SVM algorithm, which we visualize using t-SNE on 1,000 random patches per slide and color based on true tissue class and MorphoITH pseudo color. Note: in example B there is significant heterogeneity within the tumor class.

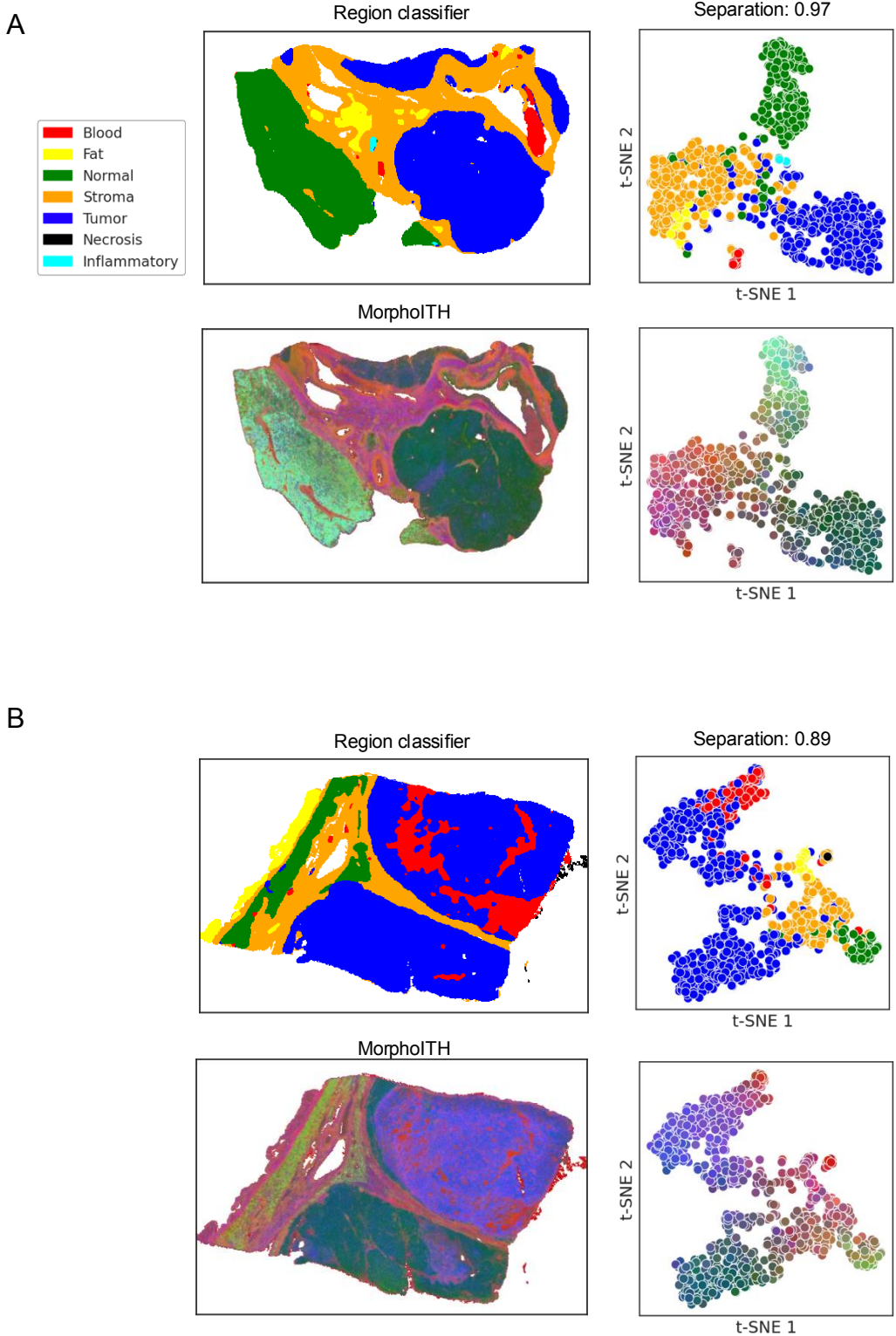

**Supplementary figure 4.** Examples of A) full, B) partial, and C) no separation of tumor vs non-tumor areas in TCGA KIRC based on MorphoITH-derived feature vectors. We show input H&E images (1st column) with their MorphoITH output in form of pseudo-colors that indicate morphological similarity (2nd column), as well as "Region Classifier" output and its simplification of tumor vs non-tumor tissue classes (3rd and 4th columns, respectively). In A), we provide legend to read region classifier output (3rd column) and its categorization into tumor vs non-tumor regions (4th column).

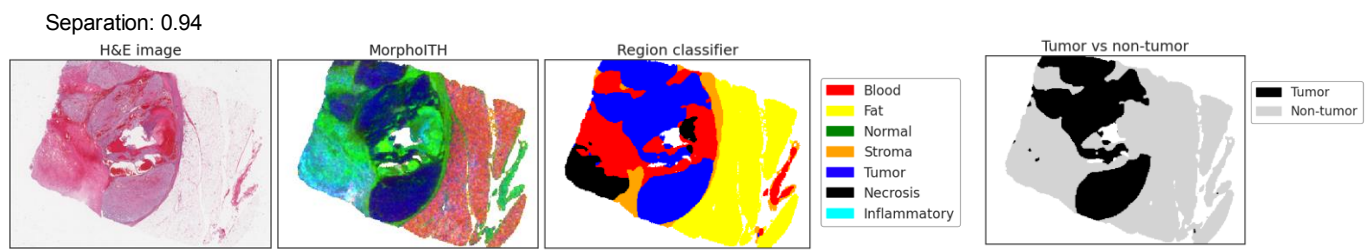

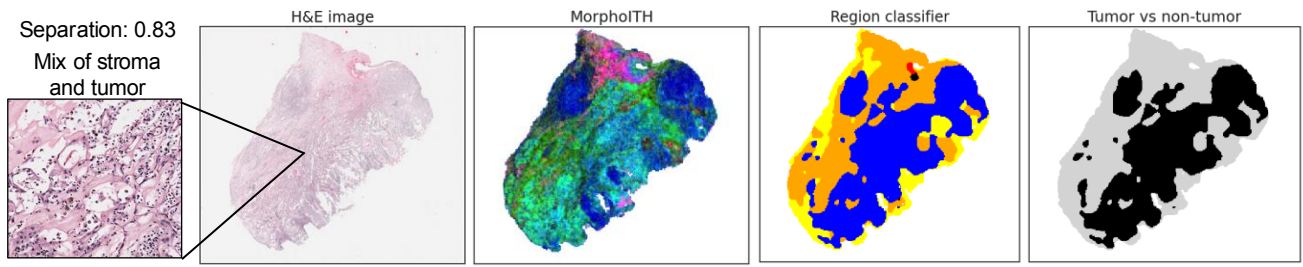

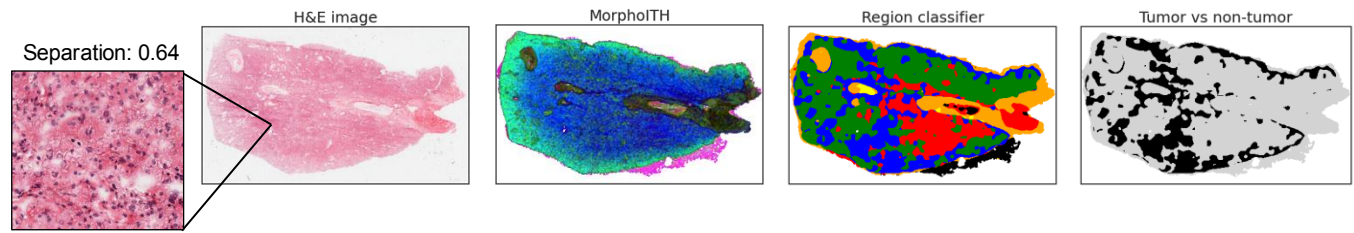

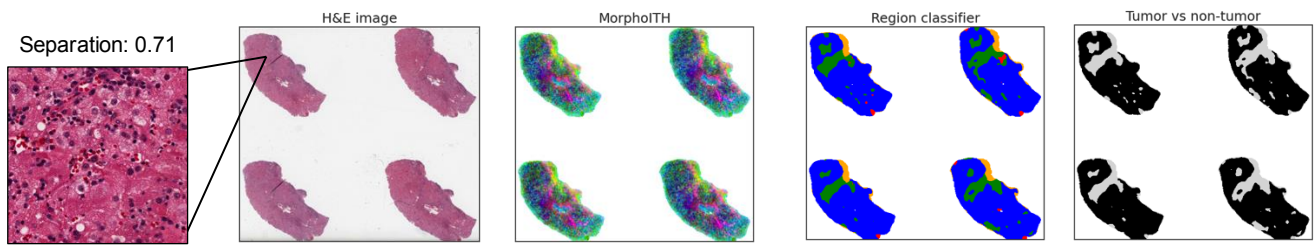

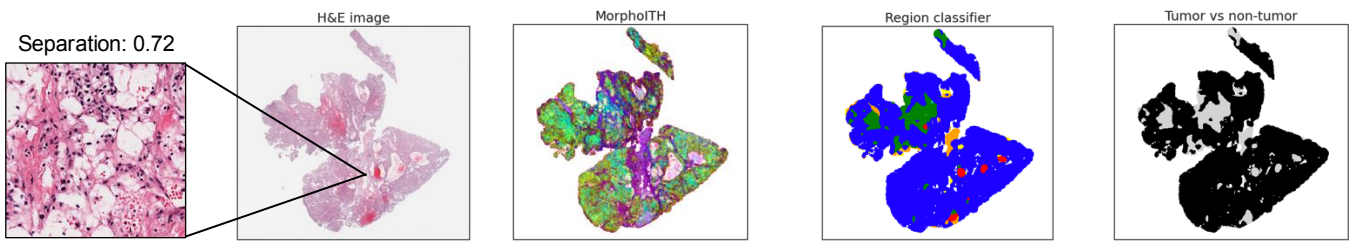

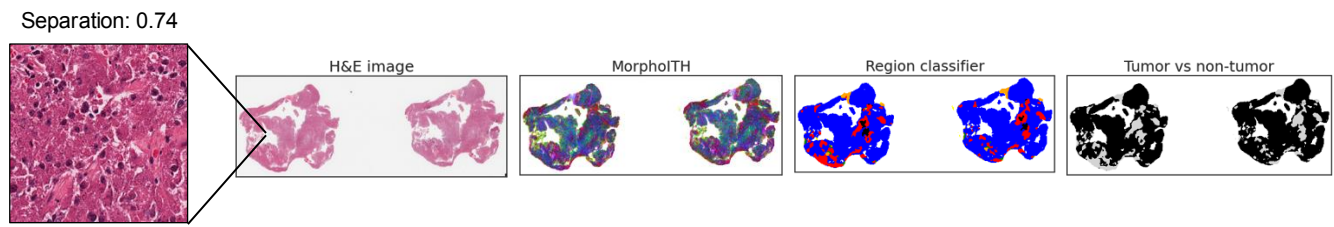

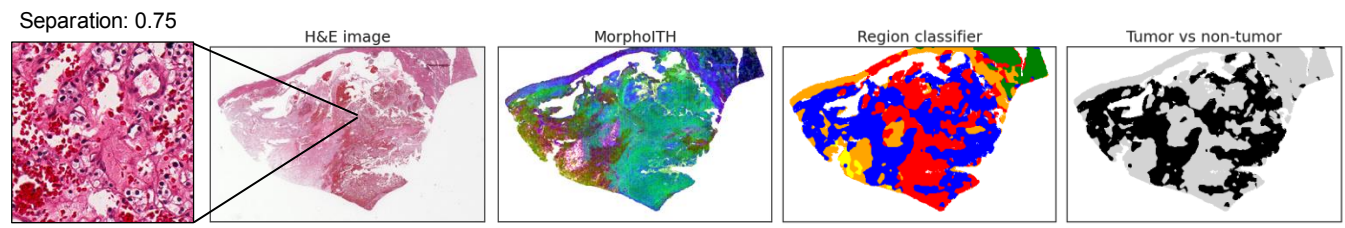

**Supplementary figure 5.** Distinct classes of vascular architectures and nuclear grade show separation based on MorphoITH similarity measure, with expectations of grade 1 vs 2, and small vs large nest architectures. Separation measure (y-axis) of A) vascular architectures and B) nuclear grades. X-axis show pairs of architectures/grades: each point in a plot corresponds to one pair of within-slide regions that were compared. Corresponding contiguity baselines and negative controls are indicated in grey. In A, different point shapes indicate whether there was a nuclear grade change between the pair that already differs in architecture.

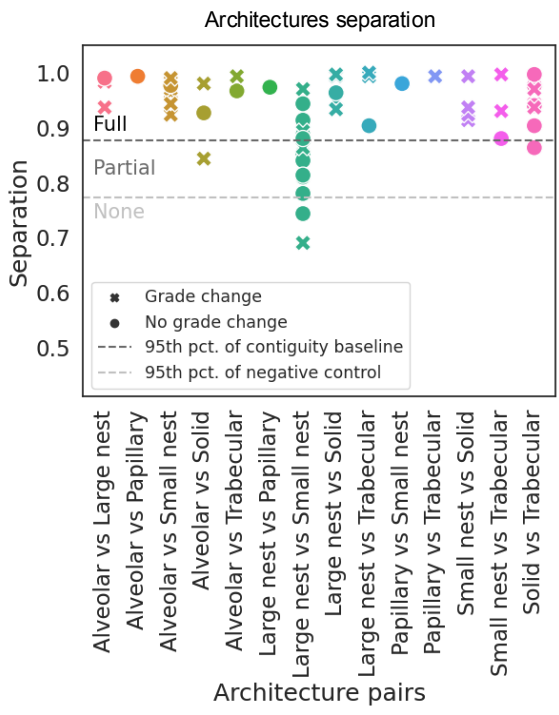
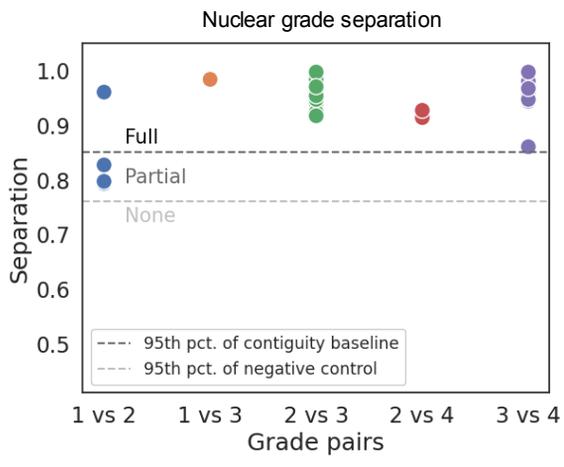

**Supplementary figure 6.** MorphoITH encompasses a broad range of morphological descriptors. A) Examples of different morphological descriptors heatmaps (nucleus size, vasculature density, eosin intensity) as compared to our measure of similarity, with output visualization pseudo-colors indicating morphological similarity. B) While MorphoITH can recognize three distinct areas A, B, and C, the individual descriptors might be undistinguishable between them. The sampled areas are indicated in the MorphoITH map with "x", and span 25 adjacent patches, with each dot in the box-plots describing mean feature value within a single patch. C) Relationships between most similar patches selected based on different descriptors. For the same slide as shown in A), we calculated 100 pairs of patches that were most similar based on the "selection features" (y-axis; morphological similarity, vasculature density, eosin intensity, random), and calculated their scaled rank (x-axis, most similar/dissimilar pair have rank 0/100, respectively) based on difference in "comparison feature" (nucleus size). "Random" row describes difference in nucleus size if patches were chosen randomly. D) The same analysis as in C), averaged cross all slides and calculated using all combinations of selection features (y-axis) and comparison features (x-axis, as opposed to just nucleus size in C). While in general the ranks of the nearest neighbors for one feature were no better than a random ranking relative to a different feature, the MorphoITH based ranking consistently succeeded in capturing points deemed similar by all other descriptors.

A

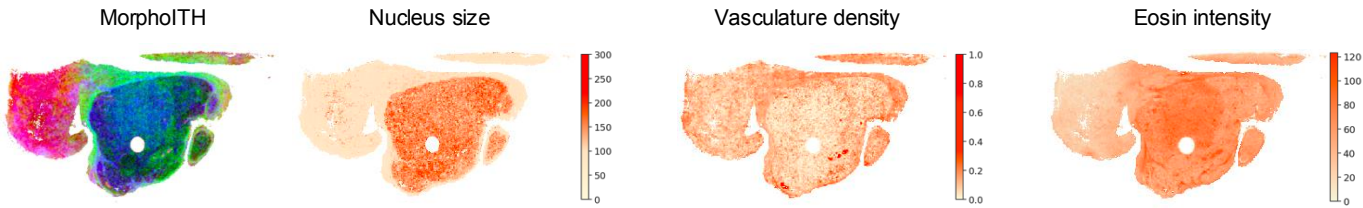

B

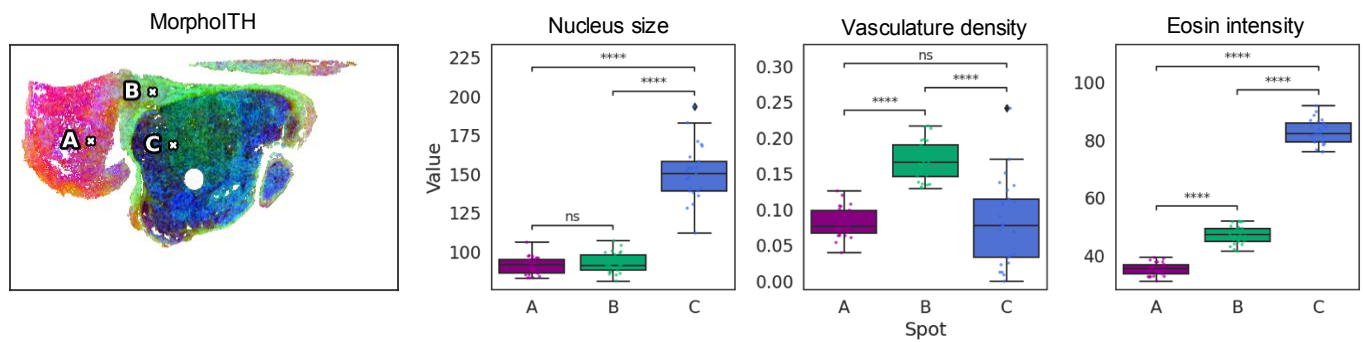

C

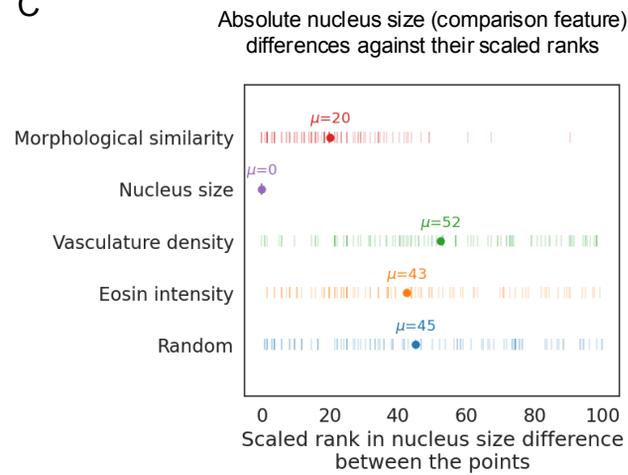

D

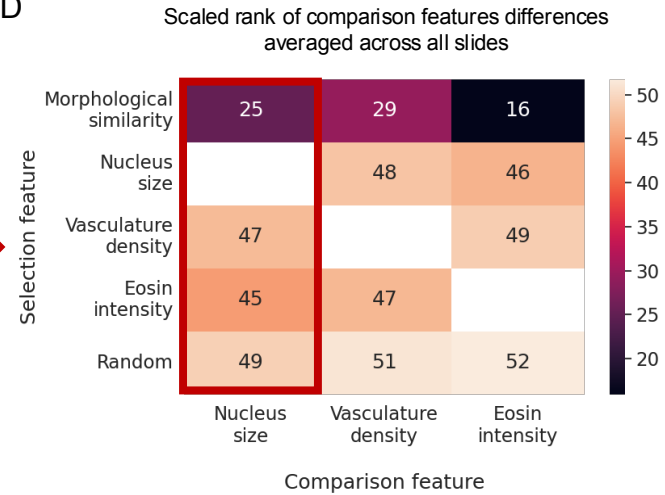

**Supplementary figure 7.** A) Examples of a) full, b) partial, and c) no separation between ground truth annotations for chosen examples of driver mutations. Left column: ground truth based on pathologist's annotations. Right column: visualization of MorphoITH output with pseudo-colors which indicate similarity in morphology. B) Overlap values (y-axis) between loss region and morphological clusters for all three driver mutations: *BAP1*, *SETD2*, and *PBRM1*, with corresponding contiguity baseline and negative control. C) Distribution of cosine distances (heterogeneity scores; x-axis) within TMA cores (blue), and between a pair of clusters within a slide which have the highest cosine distance compared to other pairs (green). On the left, examples of TMA cores with low and high heterogeneity scores. D) Focal cases (with both WT and loss in the same slide) of *BAP1* and *SETD2* have significantly higher morphological heterogeneity than slides without heterogeneity in driver mutation status. Morphological heterogeneity scores as calculated by cosine distance between the most distinct clusters within a slide (y-axis).

A
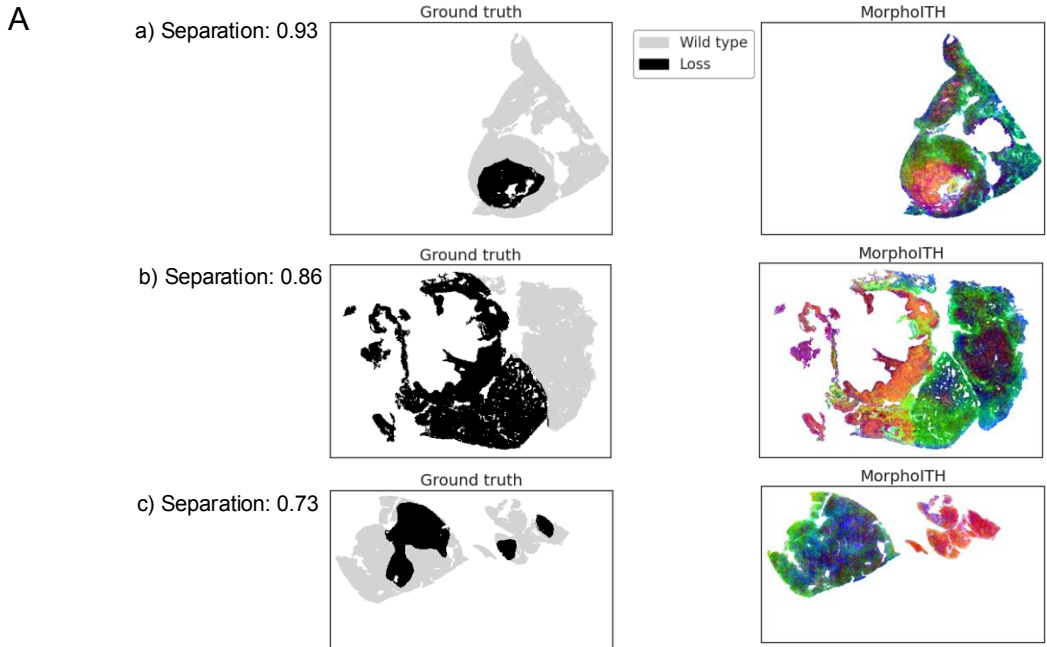

B
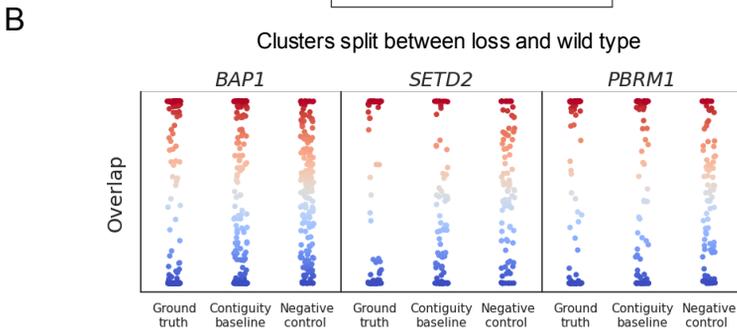

C
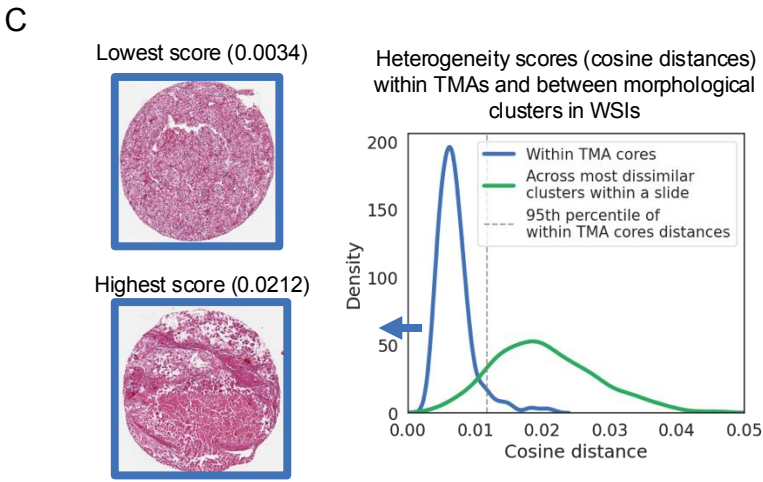

D
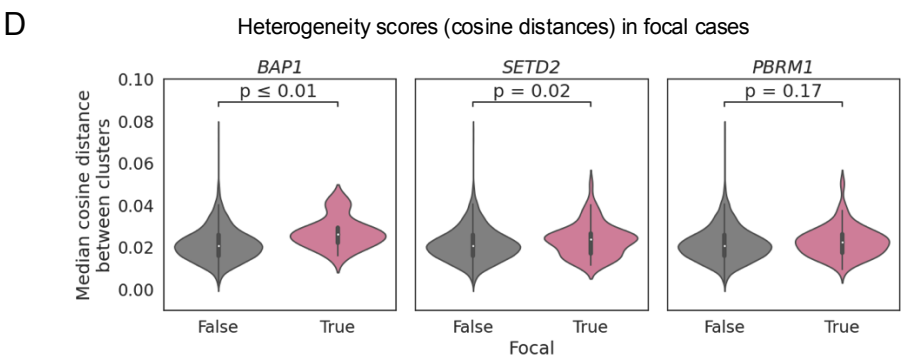

**Supplementary figure 8.** A-C) Phylogenetic trees as in Fig. 4 in phylogram form with maintained edge lengths.

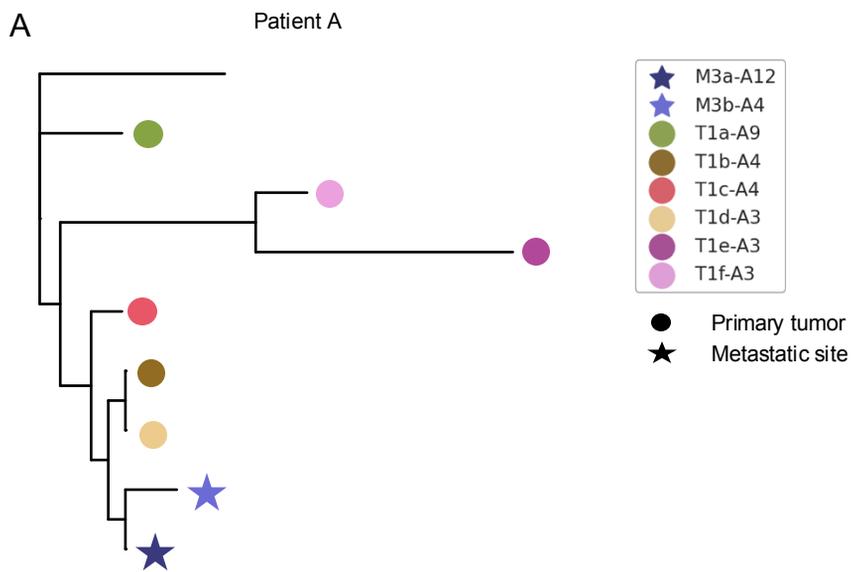

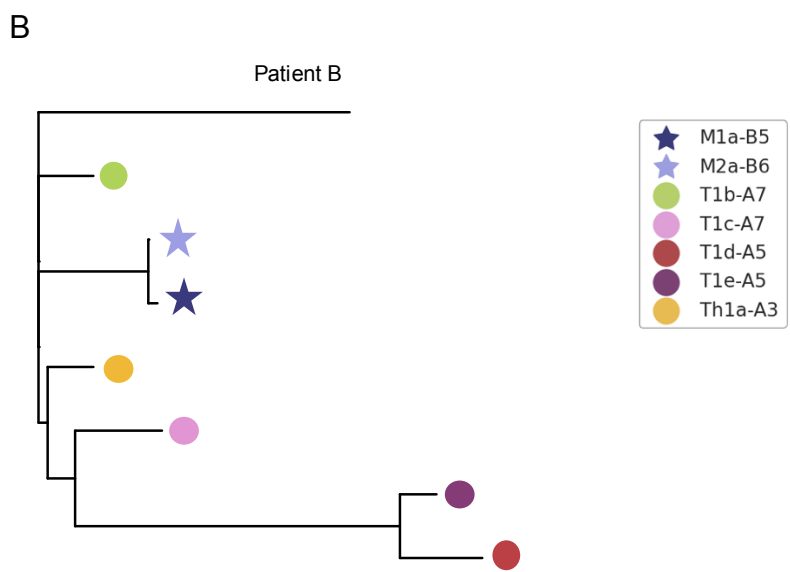

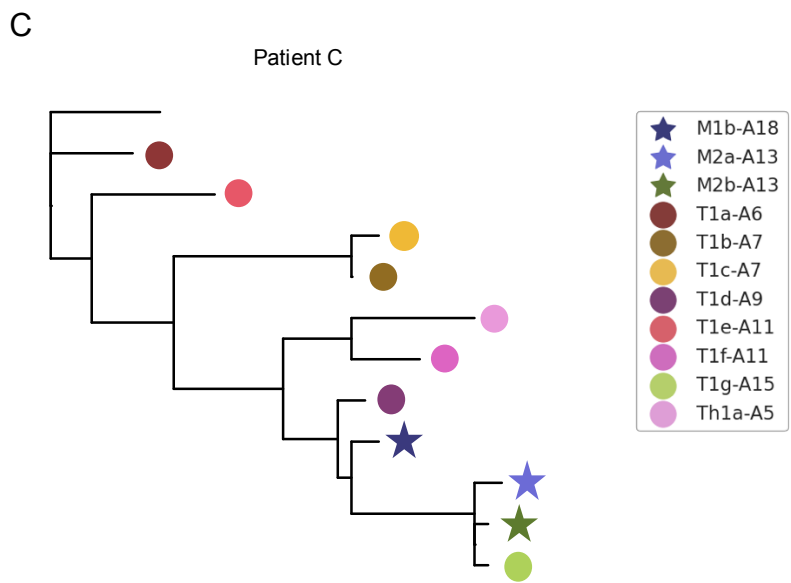

**Supplementary figure 9.** A-C) t-SNE plots of morphological similarity with overlayed patches from which the MorphoITH feature space was derived (1st column). Heatmaps of MorphoITH distances within a patient (2nd column). Mean feature representation per sample were compared using cosine distance to obtain MorphoITH distances.

A
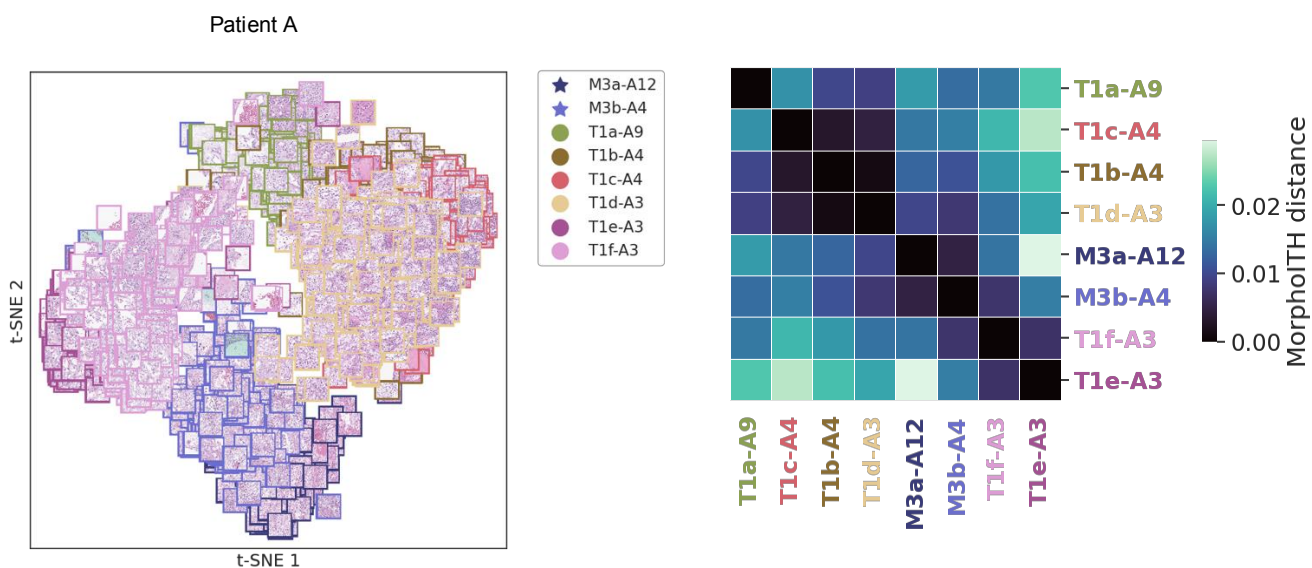

B
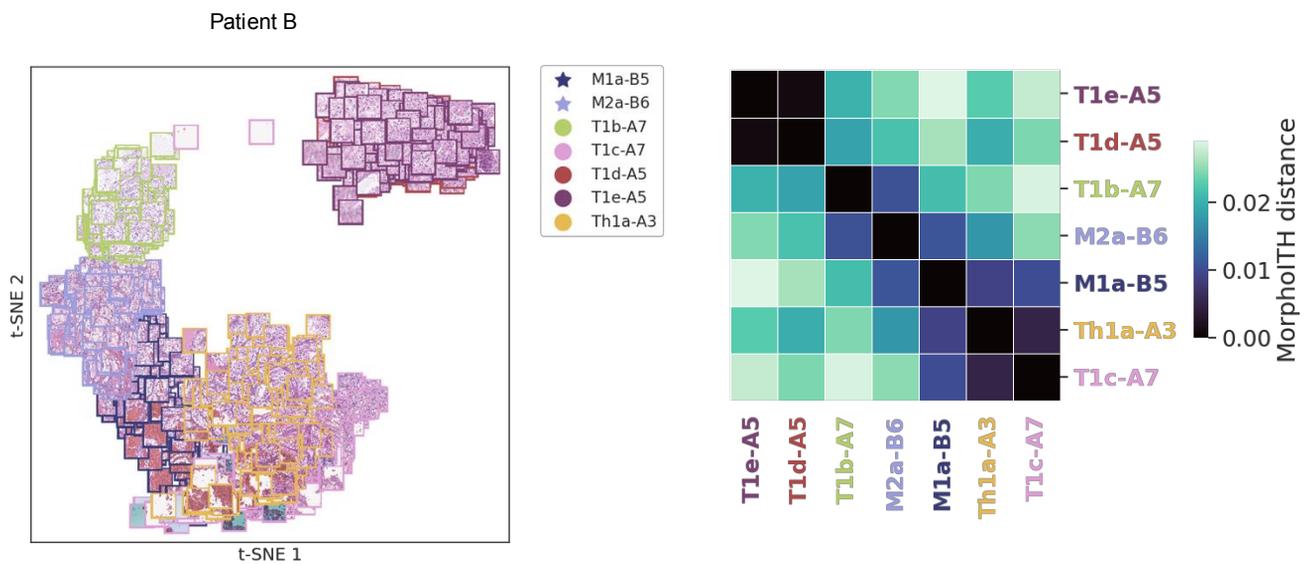

C
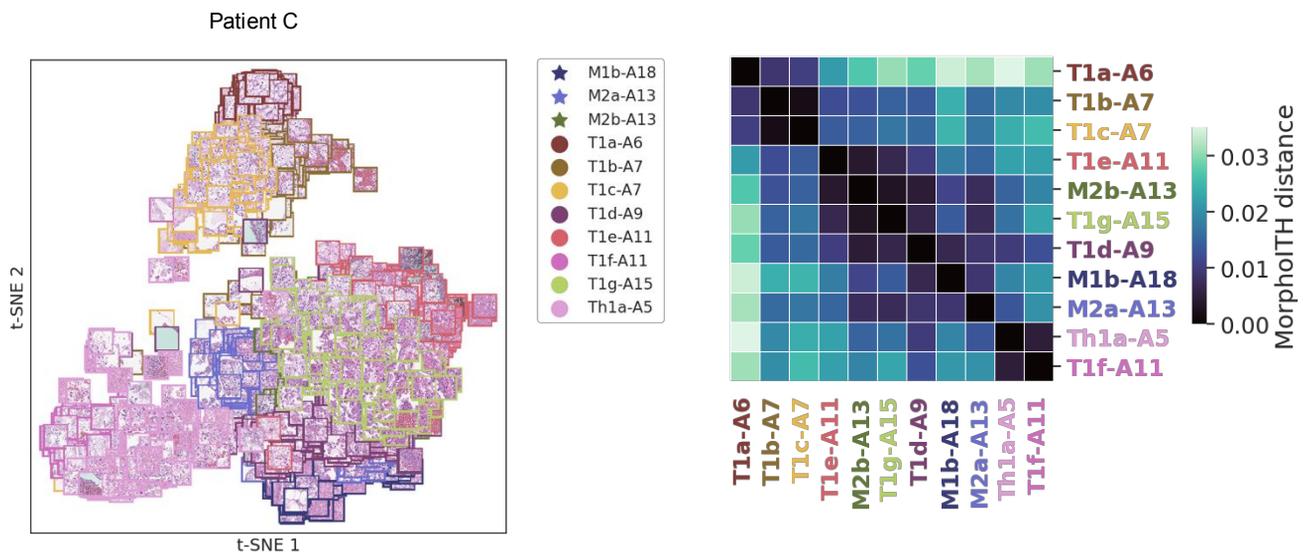

**Supplementary figure 10.** A-C) t-SNE plots of morphological similarity (each point represents one patch) and phylogenetic trees for patients A-C with overlayed nuclear grade, vascular architectures annotations, as well as H&E slide of origin.

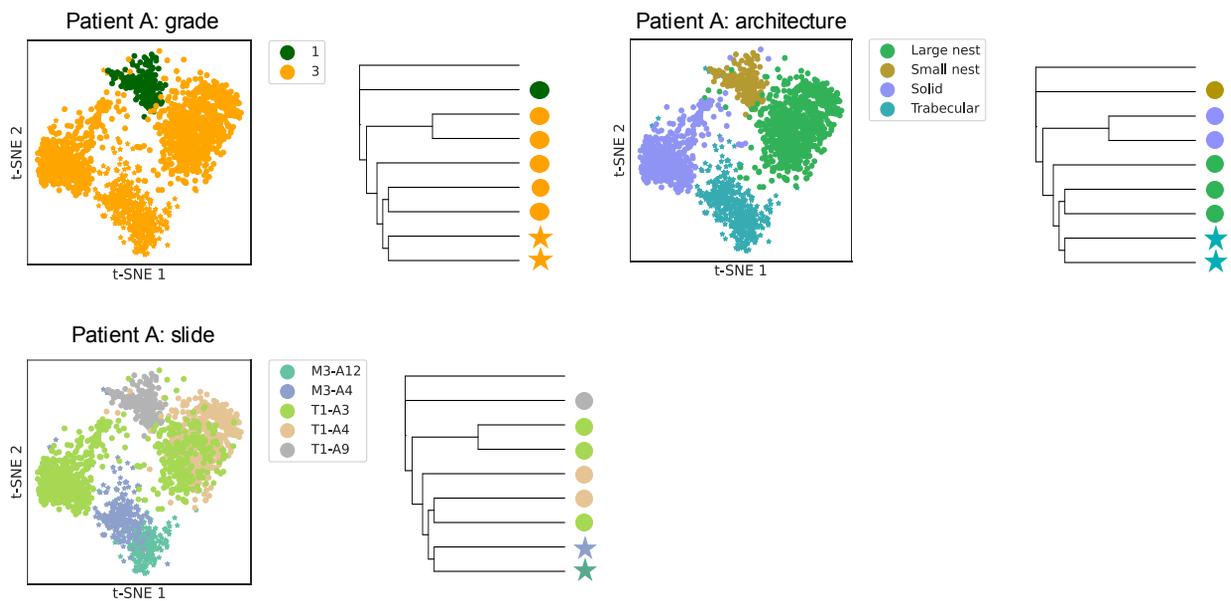

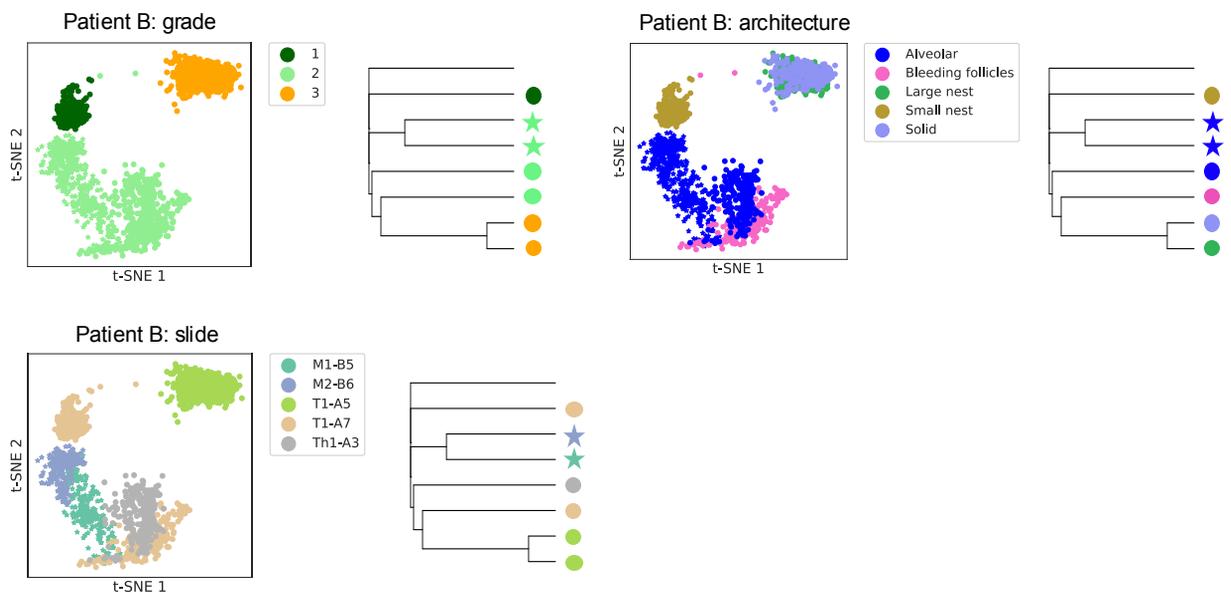

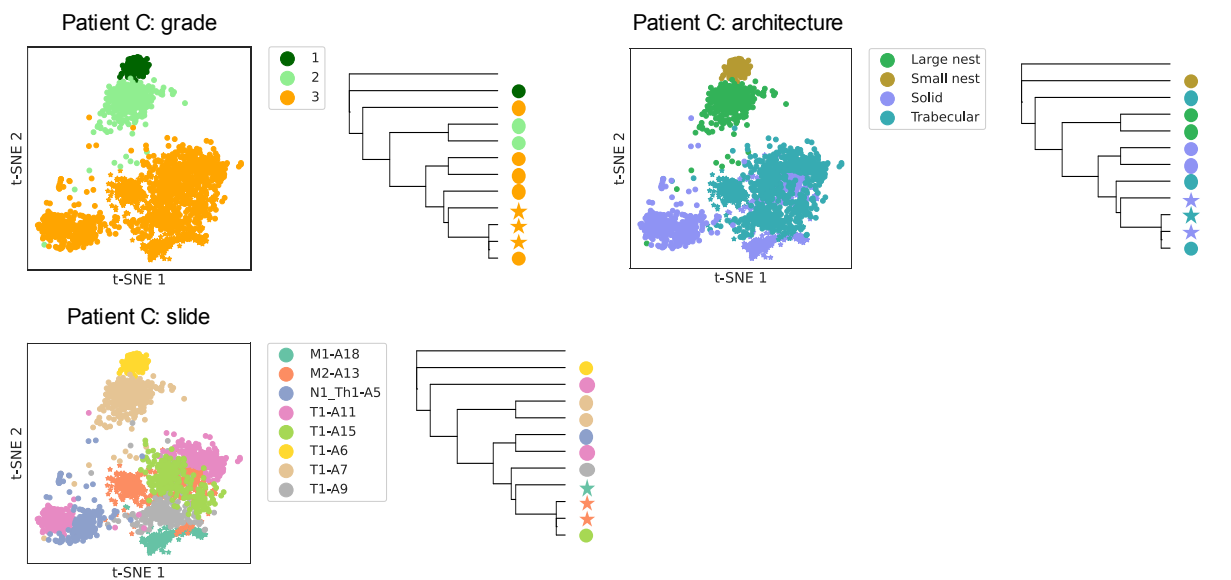

**Supplementary figure 11.** A-C) Separation measure between all samples within patients A-C (1st column), with phylogenetic trees with highlighted clades containing samples that have low separation in morphology (lower half of the separation range, up to 3 samples; 2nd column).

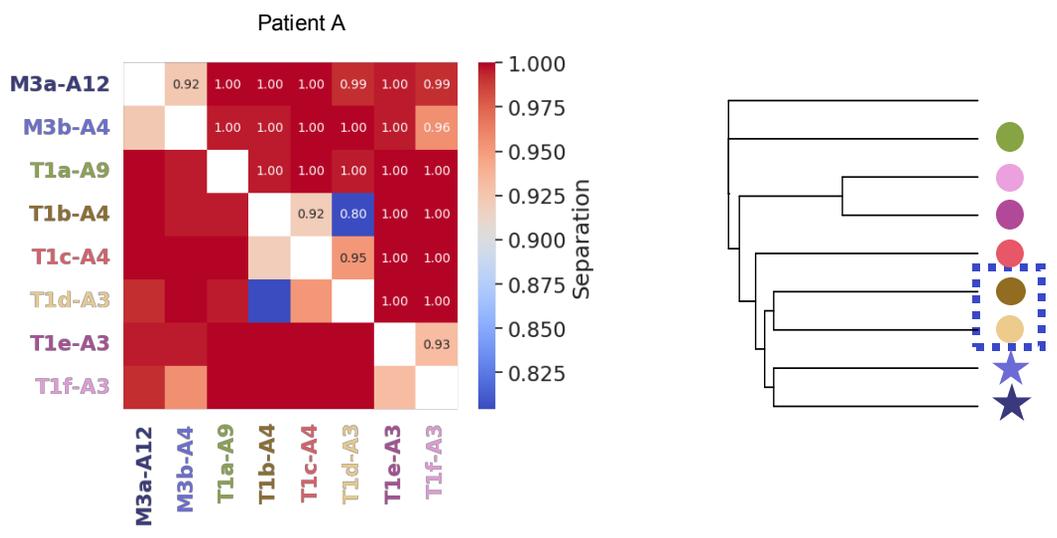

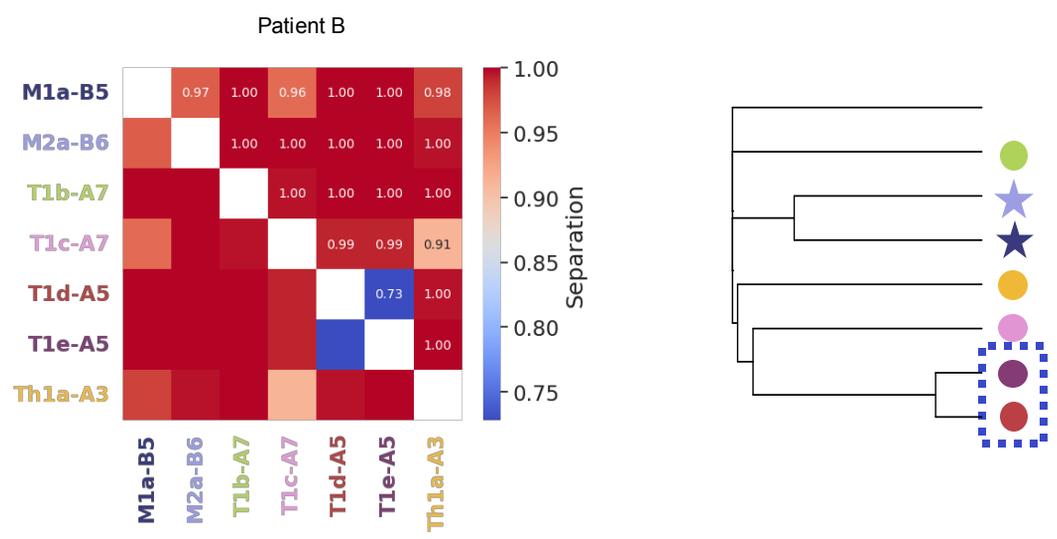

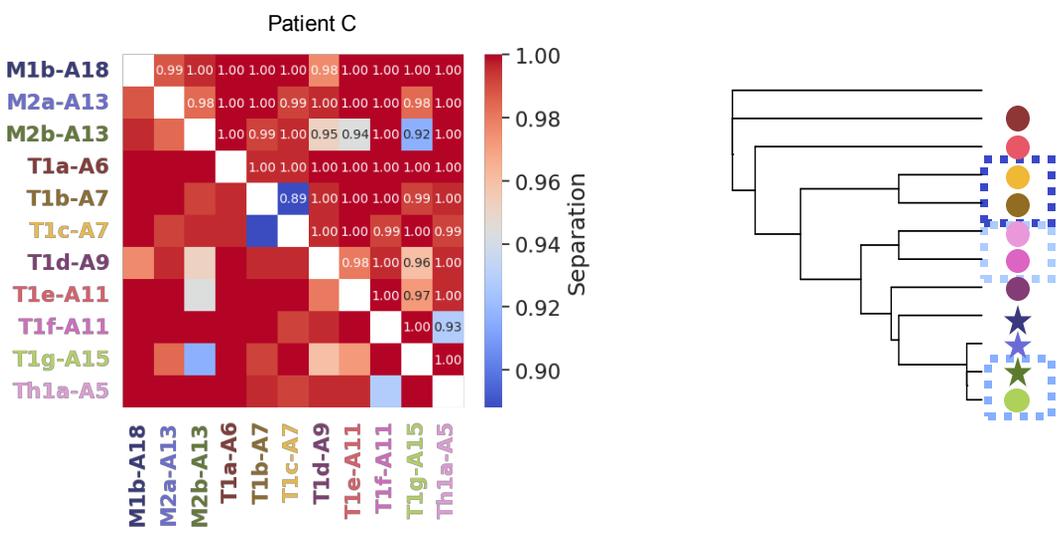

**Supplementary figure 12.** Examples of outliers from the plot show correlation between MorphoITH (x-axis) and genetic (y-axis) distances from Fig. 4D. On the left, we show examples of sample pairs that are considered similar morphologically and less so genetically (T1e-A11 vs [M2b-A13, T1g-A15]). On the right, we show examples of sample pairs that are considered different morphologically but more closely related genetically (T1a-A6 vs [M2a-A13, Th1a-A5, T1g-A15, T1d-A9, M2b-A13, M1b-A18, T1f-A11], T1c-A7 vs [Th1a-A5, M1b-A18, T1f-A11], T1b-A7 vs M1b-A8).

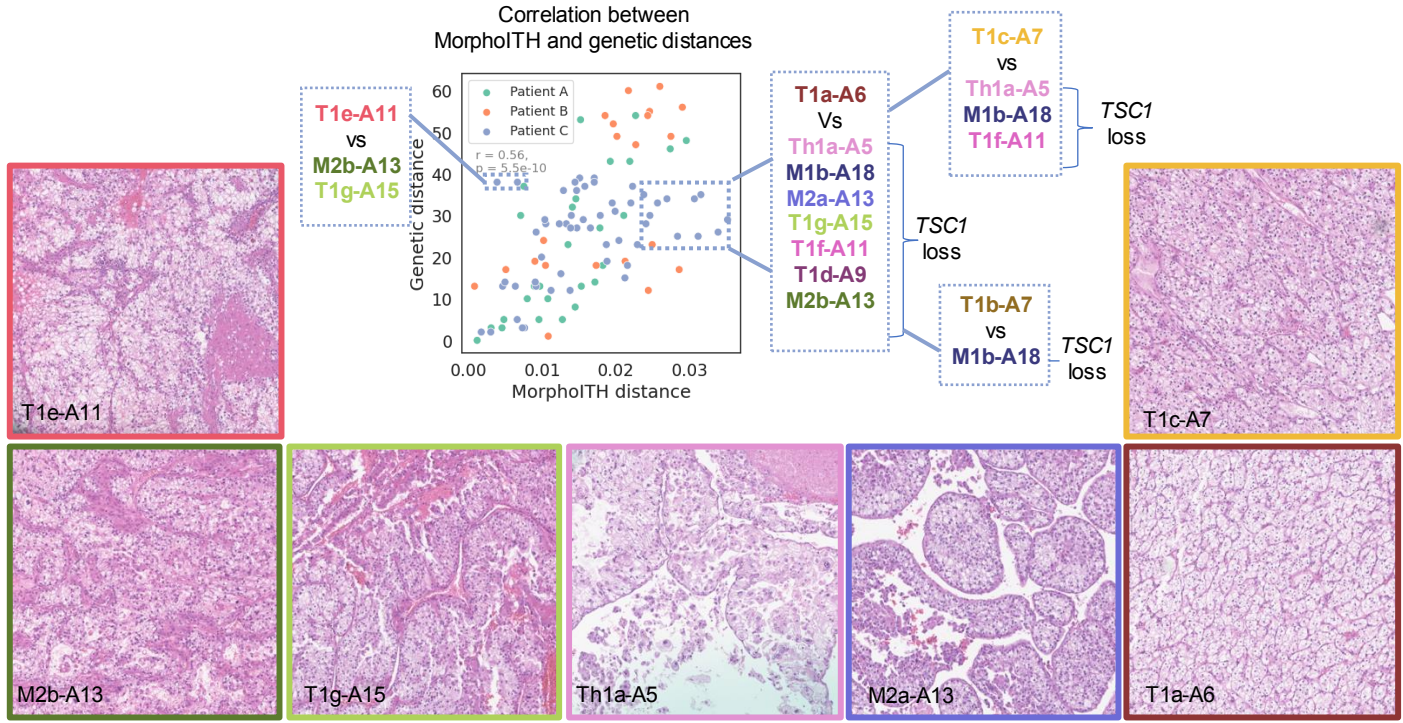

**Supplementary figure 13.** Per-patient correlation between MorphoITH and genetic distances on: A) original patches, or patches with applied normalization scheme as following: B) Macenko, C) Vahadane. Each plot includes Pearson's correlation with its p-value.

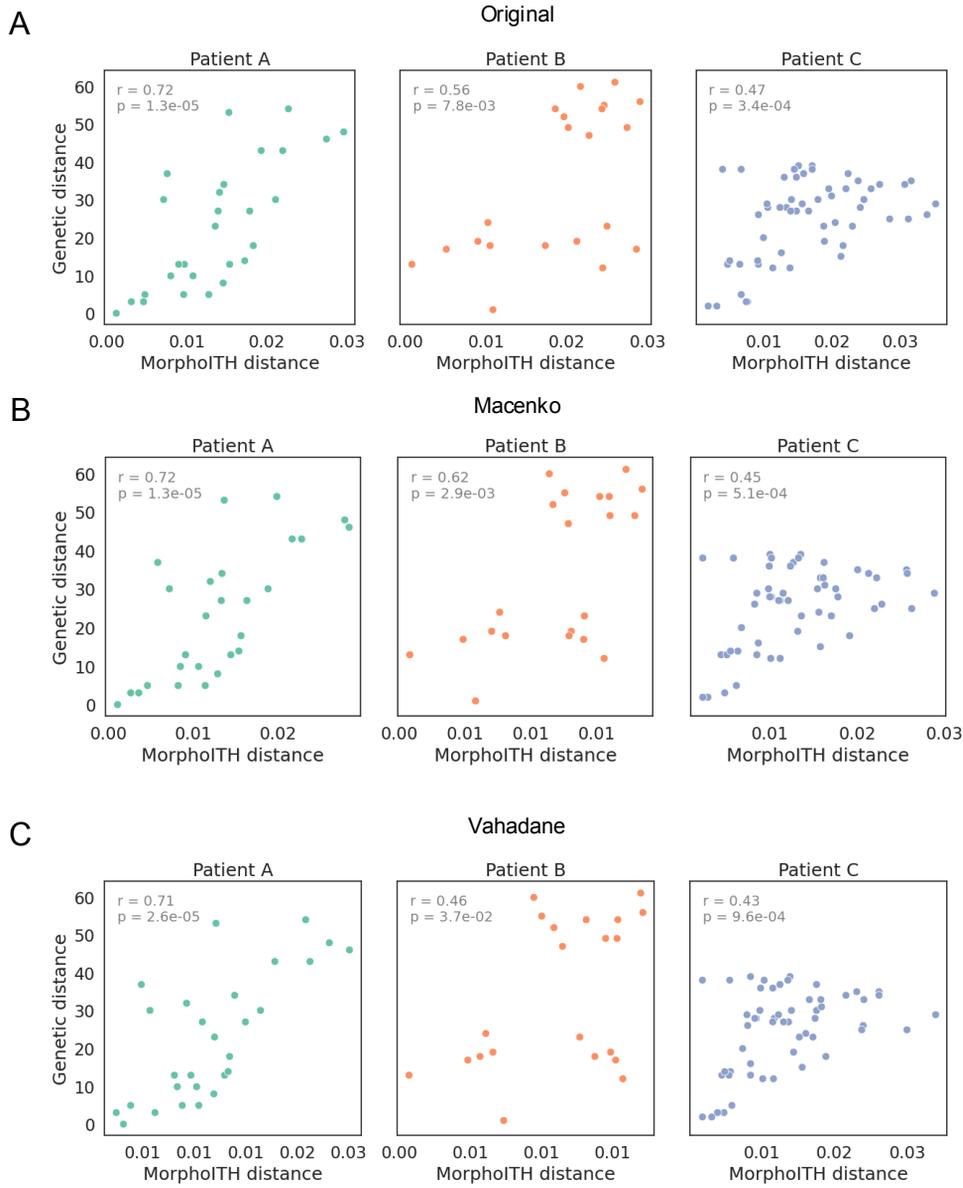